\documentclass[11pt,reqno]{amsart}

\usepackage{amsfonts,amsthm,latexsym,amsmath,amssymb,amscd,amsmath, epsf}
\usepackage[letterpaper,margin=1in]{geometry}
\usepackage{color}
\usepackage[utf8]{inputenc}
\usepackage{picture}
\usepackage{tikz}
\usepackage{graphicx}
\usepackage{verbatim}
\usepackage{tcolorbox}
\tcbuselibrary{skins}
\usepackage{booktabs,multirow}
\usepackage{siunitx}
\usepackage{etoolbox}
\usepackage{wrapfig}
\usepackage{listings}
\usepackage[foot]{amsaddr}
\usepackage{xcolor}
\definecolor{greenD}{rgb}{0.13, 0.55, 0.13}

\makeatletter
\AtBeginEnvironment{minted}{\dontdofcolorbox}
\def\dontdofcolorbox{\renewcommand\fcolorbox[4][]{##4}}
\makeatother

\usepackage[super]{nth}
\usepackage[ruled,linesnumbered]{algorithm2e}
\usepackage{algpseudocode}

\newcommand{\enVert}[2][-1]{
	\ensuremath{\mathinner{
			\ifthenelse{\equal{#1}{-1}}{ 
				\!\left\lVert#2\right\rVert}{}
			\ifthenelse{\equal{#1}{0}}{ 
				\lVert#2\rVert}{}
			\ifthenelse{\equal{#1}{1}}{ 
				\!\bigl\lVert#2\bigr\rVert}{}
			\ifthenelse{\equal{#1}{2}}{ 
				\!\Bigl\lVert#2\Bigr\rVert}{}
			\ifthenelse{\equal{#1}{3}}{ 
				\!\biggl\lVert#2\biggr\rVert}{}
			\ifthenelse{\equal{#1}{4}}{ 
				\!\Biggl\lVert#2\Biggr\rVert}{}
	}} 
}
\let\norm=\enVert
\DeclareMathAlphabet{\mathdutchcal}{U}{dutchcal}{m}{n}
\SetMathAlphabet{\mathdutchcal}{bold}{U}{dutchcal}{b}{n}
\newcommand{\set}[1]{\mathdutchcal{#1}}
 \newtheorem{theorem}{Theorem}[section]
 \newtheorem{corollary}[theorem]{Corollary}
 
 \newtheorem{proposition}[theorem]{Proposition}
 
  \newtheorem*{question*}{\sc Question}
   
 \theoremstyle{definition}
 \newtheorem{definition}[theorem]{Definition}
 \theoremstyle{remark}
 \newtheorem{remark}[theorem]{Remark}
 \theoremstyle{definition}
 \newtheorem{example}[theorem]{Example}

\theoremstyle{plain}
\newtheorem*{claim*}{Claim}

 \usepackage{listings}

\newcommand{\numberset}{\mathbb}
\newcommand{\ort}{\numberset{O}}
\newcommand{\K}{\numberset{C}}
\newcommand{\R}{\numberset{R}}

\newcommand{\M}{\numberset{M}}
\newcommand{\N}{\numberset{N}}

\newcommand{\E}{\numberset{E}}
\newcommand{\V}{\numberset{V}\text{ar}}

\usepackage{subfig}
\usepackage{float}
\usepackage{pgfplots,relsize,siunitx}
\pgfplotsset{compat=1.14}
\usepackage{tikz}
\usetikzlibrary{calc, positioning}
\usetikzlibrary{matrix}
\usetikzlibrary{patterns}

\pgfplotsset{grid style={line width=0.05pt, gray}}
\pgfplotsset{minor grid style={gray}}
\pgfplotsset{major grid style={gray}}

\title{High Order Singular Value Decomposition for plant biodiversity estimation}
\author[A. Bernardi]{Alessandra Bernardi$^\star$}
\address{$^\star$Department of Mathematics, University of Trento, 38123 Povo (TN), Italy} 
\author[M. Iannacito]{Martina Iannacito$^\dagger$}
\address{$^\dagger$Inria Bordeaux - Sud-Ouest, 200, avenue de la Vieille Tour, 33405 Talence, France}
\author[D. Rocchini]{$^\ddagger$Duccio Rocchini}
\address{$^\ddagger$Department of Biodiversity and Molecular Ecology, Fondazione Edmund Mach, Research and Innovation Centre, 38010 S. Micehle all’Adige (TN), Italy}
\email{alessandra.bernardi@unitn.it, martina.iannacito@inria.fr, duccio.rocchini@unitn.it}
\date{September 2019}

\begin{document}

\begin{abstract}
We propose a new method to estimate plant biodiversity with Rényi and Rao indexes through the so called High Order Singular Value Decomposition (HOSVD) of tensors. Starting from NASA multispectral images we evaluate biodiversity and we compare original biodiversity estimates with those realised via the HOSVD compression methods for big data. Our strategy turns out to be extremely powerful in terms of storage memory and precision of the outcome. The obtained results are so promising that we can support the efficiency of our method in the ecological framework. 

\end{abstract}

\maketitle

 \section{Introduction}\label{sec:intro}
 In order to face Earth changes in space and time, satellite imageries are nowadays being used to provide timely information over the whole globe.
 
 From this point of view, starting from the early '70s, technological improvements in remote sensors led to a growth in the number and in the size of available data. For example to store a band of the entire Earth's surface from the MODIS sensor with a low spectral resolution, $\SI{5600}{\meter}$, we need around $\SI{99}{\mega\byte}$. Therefore for rasters with higher spectral, spatial or radiometric resolutions the memory request increases significantly. 
 One may point out that $\SI{99}{\mega\byte}$ are not much, if compared with the features of modern machines. However the memory needed to process thousands of images might represent a crucial issue.
 Moreover, in order to perform any index computation over these images, they should be loaded into the computer RAM, which usually has lower capacity and is occupied also by the system application and by our computing script.\\ 
 Since the advent of big data, which include multispectral and hyperspectral images, improving storage and analytical techniques became fundamental. Mathematicians are facing this challenge together with computer scientists, physicists and engineers (cf. e.g. \cite{mate0, mate13,ACAR201141,LIM2010311,10.21468/SciPostPhysLectNotes.8,Bini_1994, Benzi_2016, BERNARDI201778,Rizzi,Ver, MR2895192, Ballico_2019}). The usual structure for storing multispectral images are \emph{tensors}.
 For matrices approximation  there exists an optimal technique, called \emph{Singular Values Decomposition}, SVD (cf. \cite{Schmidt,Steward,doi:10.1002/mana.19590200306,EY}). One application of SVD is image compression (cf. e.g. \cite{WANG201563, Jia_2012, GoluVanl96}). Since we need to store data in tensors, the generalisation of matrices, we will present \emph{High Order Singular Value Decomposition} (HOSVD) firstly introduced in \cite{mate0, mate13}, a generalisation to tensors of SVD to tensors. There are various techniques to approximate tensors by taking advantage of the HOSVD, those that will be implemented in this paper are the so called Truncated HOSVD (T-HOSVD) and Sequentially Truncated HOSVD (ST-HOSVD). Even if only for some special cases HOSVD provides optimal results (\cite{mate10, BDHR, Draisma2018}), it is possible to present an estimate of the tensor approximation errors. Indeed the core of the present paper will be application of T-HOSVD and ST-HOSVD techniques, their modern versions and the error made by using them. Indeed in the last section we will apply some possible HOSVD implementations to tensors in which we stored RED and NIR bands. Next from the compressed tensors we get NDVI rasters and over them we compute biodiversity indexes. So we will be able to compare our biodiversity estimations from compressed data with those realised over original data. \\
 As far as we know, {this is the first attempt to estimate biodiversity with the presented indexes from HOSVD compressed data}. Moreover the obtained results are extremely promising, letting us support its efficiency in ecological framework. 
 \normalcolor
\section{Ecological background}\label{sec:eco}
One of the main component of Erath's biosphere is biodiversity, which is in strict relationship with the planet status in space and time. Measuring biodiversity over wide spatial scales is difficult concerning time, costs and logistical issues, e.g.: (i) the number of sampling units to be investigated, (ii) the choice of the sampling design, (iii) the need to clearly define the statistical population, (iv) the need for an operational definition of a species community, etc. (cf.~\cite{chiarucci_2007}). Hence, ecological proxies of species diversity are important for developing effective management strategies and conservation plans worldwide (cf.~\cite{rocchini_2010}). From this point of view, environmental heterogeneity is considered to be one of the main factors associated to a high degree of biological diversity given that areas with higher environmental heterogeneity can host more species due to the greater number of available niches within them. Therefore, measuring heterogeneity from satellite images might represent a powerful approach to gather information of diversity changes in space and time in an effective manner (cf.~\cite{zellweger_2019}).
Nowadays, the advent of satellites has made possible having real images of a territory, even with a remarkable quality. This approach is behind the the remote sensing discipline. Many definitions have been proposed during the years for this subject. We refer to the one presented in~\cite[~p.6]{eco8}.
\begin{definition}
	Remote sensing is the practice of deriving information about Earth's land and water surfaces using images acquired from an overhead perspective, using electromagnetic radiation in one or more regions of the electromagnetic spectrum, reflected or emitted from the Earth's surface.	
\end{definition}
This field of study is based on a well known physical phenomenon: different materials and different organisms absorb and reflect electromagnetic radiations differently.
Most of the modern satellite sensors can acquire multiple images, divided in the so called bands.
\begin{definition}
	\emph{Bands} or \emph{channels} are the recorded spectral measurements.
\end{definition} 
Depending on the sensor features, we can acquire images dived into the different bands.
\begin{definition}
	\emph{Multi-spectral sensor} can acquire from 4 up to 10 bands. \emph{Hyper-spectral sensor} can acquire more than 100 bands. 
\end{definition}
Data used in the present paper come from the \emph{MODerate-Resolution Imaging Spectroradiometer} sensor, or MODIS, built on both the satellites Terra and Aqua, cf.~\cite{mod1}. MODIS measures 36 bands in the visible and infrared spectrum, at different resolutions. Firstly there are RED and NIR bands with pixel size of 250km and next 5 bands, still with RED and NIR, at 500m of spatial resolution. These are extremely useful for land observation. The remaining bands with 1km of resolution consist of monitoring images from visible spectrum, MIR, NIR and TIR. The data are registered four times a day: twice daytime and twice night-time. Images usually include the entire Earth's surface.\\
From the 60s the scientific community highlighted two important relations between spectral measurements and biomass:
\begin{itemize}
	\item a direct relation between NIR region and biomass, i.e. greater the biomass greater the NIR reflected radiation measured and vice versa;
	\item an inverse relation between RED spectral region and biomass, i.e. greater the biomass, lower the RED visible reflected spectrum measured and vice versa.
\end{itemize} 
Therefore the relation between NIR and RED reflectance is central for the vegetal biomass estimation. The aim of vegetation indexes is measuring biomass or vegetative vigour on the base of digital brightness values. The one used in the present article is the \emph{Normalized Difference Vegetation Index}, or NDVI, presented in 1974 by Rouse and others, cf.~\cite{eco17}. 
\begin{definition}
	\label{def2_3_1}
	Given a region $R$, let $RED,NIR\in\M^{m \times n}(\R)$ be respectively the RED and the NIR raster band of $R$ imagery. The	\emph{normalized difference vegetation index} of region $R$ is $NDVI\in\M^{m\times n}(\R)$ such that 
	\[\text{NDVI}_{ij}=\dfrac{NIR_{ij} - RED_{ij}}{NIR_{ij}+RED_{ij}}
	\]
	for every $i\in\{1,\dots,m\}$ and for every $j\in\{1,\dots,n\}$, when it is defined.
\end{definition}
The innovative idea of ~\cite{eco31} was the application of information theory studies and ecological indexes to remote sensed images. The ecological indexes used in the present paper are the Rao and Rényi ones, which we are going to define.
\begin{definition}
	Given a spectral image of a sample area, let $N$ be the image radiometric resolution and let $p_i$ be the relative abundances of the $i$-th value for every $i\in\{1,\dots,N\}$.
	Fixed a distance function $d$, we build up a pairwise spectral difference matrix $D\in\M^{N}(\R)$ such that
	\[D_{ij}=d(i,j)
	\]
	for every $i,j\in\{1,\dots,N\}$. The \emph{Rao's Q index} for the sample area is
	\[I_{RQ}=\sum_{j=1}^N\sum_{i=1}^Np_ip_{j}D_{ij}.
	\] 
\end{definition}
\par \begin{definition}
	Given a sample area with $N$ species and defined $p_i$ the relative abundances for every $i\in\{1,\dots,N\}$, in decreasing order, the \emph{Rényi index} is
	\[I_{R}=-\log\sum_{i=1}^Np_i^2.
	\] 
\end{definition}
The main difference between them is that Rao index takes into account both the frequencies and the numerical values of each pixel. On the other side Rényi index considers only pixels frequencies. Lastly notice that rasters usually are split into small chunks, called \emph{windows}, over which the biodiversity chosen index is computed. We need to generate Rényi's and Rao's computation codes to measure how much biodiversity information is lost using approximated tensor data. Since we keep the ecologists moving window approach, our implementation presents also a very basic multi-core modality. The multi-cores Rény's and Rao's implementations crux is presented in the Appendix, section \ref{app:1}.   

\section{Mathematical background}\label{sec:mate}
The celebrated Schmidt \cite{Schmidt,Steward}, Mirsky \cite{doi:10.1002/mana.19590200306}, Eckhart, Young \cite{EY} theorem states that for every rank-$r$ matrix $A\in\M^{m\times n}(\K)$  there exists the best rank $s$ approximation for every $s< r$ in the Frobenius norm, actually the result holds for arbitrary ($O_m\times O_n$)-invariant norms \cite{Mir2}.
The famous generalization of this result \cite{mate0} to higher order tensors  fails in general to produce an output that is the ``best" approximation of a given tensor, cf. \cite{mate13}. It succeeds for the so called orthogonally decomposable tensors \cite{BDHR}. Another type of generalization of the concept of ``best rank-$r$ approximation of a tensor" proposed in \cite{Draisma2018} works for general tensors.
Despite the possible non-existence of the best approximation of a tensor under the construction of \cite{mate0}, this technique turns out to be very convenient from the computational point of view and it is possible to prove that the outcome is a ``quasi-optimal solution.
In this paper we will mainly use the technique popularized by L. De Lathauwer and J. Vandewalle in \cite{mate0}, the so called   \emph{High Order Singular Value Decomposition} (HOSVD), we will study two types of errors made by the specific approximation, we will make a comparison among them and we will show that they will be small enough to have very precise results. 

\medskip

Here some basic mathematical preliminaries. The number field will always be the complex field of numbers $\mathbb{C}$.
\begin{definition}
	Let $V$ be a linear subspace of the tensor space $\K^{n_1}\otimes\dots\otimes\K^{n_d}$. If for every $i\in\{1,\dots,d\}$, there exist a subspace $V_{i}\subseteq \K^{n_i}$ such that 
	\[V=V_1\otimes\dots\otimes V_d,
	\] then $V$ is \emph{a separable tensor subspace} of $\K^{n_1}\otimes\dots\otimes\K^{n_d}$.
\end{definition}
Remark that not every subspace of $\K^{n_1}\otimes\dots\otimes\K^{n_d}$ is separable.
The structure of a separable tensor space has some consequences on its elements.
\begin{definition}
	The \emph{multilinear rank} of a tensor ${A}\in\K^{n_1}\otimes\dots\otimes\K^{n_d}$ is the $d$-uple $(r_1,\dots,r_d)$ with the property that $r_i$ is the minimal dimension of a subspace $V_i \subset \mathbb{C}^{n_i}$ such that $A\in V_1 \otimes \cdots \otimes V_d$
	for every $i\in\{1,\dots,d\}$.
\end{definition}
Let ${A}\in\K^{n_1}\otimes\dots\otimes\K^{n_d}$ be a tensor and let $(r_1,\dots,r_d)\in\N^d$. We will discuss if there exists, and in case how to determine, a tensor ${M}$ of multilinear rank lower or equal component-wise than $(r_1,\dots,r_d)$ which minimizes the Frobenius norm of the tensor difference, i.e.
\[ {M} = \arg\inf_{\text{mlrank}({T})\le(r_1,\dots,r_d)}\norm[1]{{A}-{T}}.
\] 
This problem is also known as \emph{Low MultiLinear Rank Approximation} (LMLRA). M. Ishteva and L. De Lathauwer firstly stated this problem and introduced this acronym, cf.~\cite{mate11}. 
Looking for a tensor ${M}\in\K^{n_1}\otimes\dots\otimes\K^{n_d}$ satisfying the stated rank properties means searching a subspace ${V}_i\subset\K^{n_i}$ of dimension $r_i$ for every $i\in\{1,\dots,d\}$ such that, if the approximation tensor ${M}$ exists, it belongs to ${V}_1\otimes\dots\otimes{V}_d$. 

The techniques of T-HOSVD and ST-HOSVD  are described  by L. De Lathauwer, B. De Moor and J. Vandewalle, in \cite{mate0} and by N. Vannieuwenhoven, R. Vandebril and K. Meerbergen,  in \cite{mate13} respectively. Here we will recall the fundamental aspects for the reader convenience. 

\begin{definition}\label{multilinear_multiplication}
    The \emph{multilinear multiplication} of an order-$d$ tensor $A\in \mathbb{C}^{n_1 \times \cdots \times n_d}$ by a $d$-uple of matrices $(M_1,\ldots,M_d)$, $M_i\in \mathbb{C}^{n_i \times m_i}$ is
\[
(M_1, \ldots, M_d) \cdot {A} := \sum_{i=1}^r (M_1 {a}^{i}_{1}) \otimes \cdots \otimes (M_d {a}^{i}_{d}),
\]
whenever ${A}=\sum_{i=1}^ra^i_1\otimes \cdots \otimes a^i_d$. The resulting tensor belongs to $\mathbb{C}^{m_1 \times \cdots \times m_d}$. 
\end{definition}


\begin{definition}[\cite{MP1,MP2,MP3}]\label{MP}
	Given a matrix $M\in\M^{m\times n}(\K)$, the \emph{Moore-Penrose inverse} 
	is a matrix
	$M^\dagger\in\M^{n\times m }$ of $M$ satisfying the following properties: 
	\begin{itemize}
		\item $MM^\dagger M =M $;
		\item $M^\dagger M M^\dagger = M^\dagger$;
		\item $(MM^\dagger)^H=MM^\dagger$;
		\item $(M^\dagger M)^H = M^\dagger M$.
	\end{itemize}
\end{definition}
If $B=(M_1, \ldots, M_d) \cdot {A}$ in Definition \ref{multilinear_multiplication} with $M_i$'s square invertible matrices, then 
\[(M^\dag_1,\dots,M^\dag_d)\cdot {B}={A}.
\]

A tensor also naturally defines the following multilinear  maps.
\begin{definition}
Let $A\in \mathbb{C}^{n_1}\otimes \cdots \otimes \mathbb{C}^{n_d}$. For any $k \in \{1, \ldots, d\}$ we can define the $k$-th \emph{standard flattening} of $A$ as     
\begin{align*}
{A}_{(k)} : (\mathbb{C}^{n_1})^* \times \cdots \times (\mathbb{C}^{n_{k-1}})^* \otimes (\mathbb{C}^{n_{k+1}})^* \otimes \cdots \otimes (\mathbb{C}^{n_{d}})^* &\to \mathbb{C}^{n_k} \\
({w}_1, \ldots, {w}_{k-1}, {w}_{k+1}, \ldots, {w}_d) &\mapsto ({w}_1, \ldots, {w}_{k-1}, I, {w}_{k+1}, \ldots, {w}_d)^T \cdot {A}.
\end{align*}

More generally, let ${p} \sqcup {q} = [1,d]$ be a partition of $d$ with $s=|{p}|$ and $t=|{q}|$. Then, we can associate  to ${A}$ even more multilinear maps, namely 
\begin{align*}
 {A}_{({p};{q})} : (\mathbb{C}^{n_{q_1}})^* \times \cdots \times (\mathbb{C}^{n_{q_t}})^* &\to \mathbb{C}^{n_{p_1}} \otimes \cdots \otimes \mathbb{C}^{n_{p_s}} \\
 ({w}_1, \ldots, {w}_t) &\mapsto {w}_1^T \cdot_{q_1} \cdots {w}_{t-1}^T \cdot_{q_{t-1}} {w}_t^T \cdot_{q_t} {A}
\end{align*}
whose $\prod_{j=1}^s n_{p_j} \times \prod_{j=1}^t n_{q_j}$ matrix is
\[
 {A}_{({p};{q})} = \sum_{i=1}^r ({a}^{i}_{p_1}\otimes\cdots\otimes {a}^{i}_{p_s})({a}^{i}_{q_1} \otimes \cdots \otimes {a}^{i}_{q_t})^T.
\]
 This interpretation of ${A}$ is called a \emph{flattening} of $A$.
\end{definition}

The punchline of the HOSVD is to apply the SVD to the flattenings of a given tensor and to use the More-Penrose transform, Definition \ref{MP}, to build an approximated tensor. 
As already outlined, this technique is optimal for the so called orthogonally decomposable (ODeCo) tensors, cf. \cite{BDHR}, while for other cases there are no evidences that this procedure would lead to ``the best multilinear-rank approximation" of a given tensor. It's worth noting that, unlike matrices, tensors of higher order can fail to have best rank-$r$ approximations, cf. \cite{VinLek}. Anyway the various algorithms of the HOSVD are extremely explicit and it is possible to estimate the measure of the error made by thanking this approximation. We will show that for the applied purpose of this paper the approximation is very good in the considered problem.



\medskip

HOSVD provides a sparse representation of the given tensor, whose costs in terms of storage use can be easily computed. Indeed given ${A}\in\K^{n_1}\otimes\dots\otimes\K^{n_d}$ whose multilinear rank is known to be $(r_1,\dots,r_d)$, from HOSVD we get that ${A}=(U_1,\dots,U_d)\cdot {C}$ with $U_i$ rank-$r_i$ orthogonal $(n_i \times r_i)$-matrices and $C\in \mathbb{C}^{r_1}\otimes \cdots \otimes \mathbb{C}^{r_d}$ the so called \emph{core tensor}. The storage of each matrix $U_i$ costs $n_ir_i$ memory units for every $i\in\{1,\dots,d\}$ and storing the core tensor ${C}$ needs $\prod_{i=1}^dr_i$ memory units. In conclusion the sparse representation of ${A}$ costs 
\[\sum_{i=1}^dn_ir_i+\prod_{i=1}^dr_i \text{ memory units}
\]
which is extremely better than $\prod_{i=1}^d n_i$.

The following proposition reveals the main strategy of the HOSVD.
\begin{proposition}[\textbf{HOSVD}~\cite{mate13}]
	\label{prop3_1}
	Let ${V}={V}_1\otimes\dots \otimes {V}_d$ be a separable tensor subspace of $\K^{n_1}\otimes\dots\otimes\K^{n_d}$ with $\dim V_i=r_i$, $i=1, \ldots , d$. Let 
	\[\set{B}=\{u^1_{i_1}\otimes\dots\otimes u^d_{i_d}\}_{j=1, \ldots , r_i}	\]
	be an orthogonal basis of ${V}$ for the standard product, and let $(U_i)=(u^i_{j_i})_{i=1, \ldots ,d; j_i=1, \ldots , r_i}$ be the corresponding orthogonal bases for the $V_i$'s, i=$1, \ldots ,d$.
	 The projector \[\pi_i:\K^{n_1}\otimes\dots\otimes\K^{n_d}\mapsto\K^{n_1}\otimes\dots\otimes\K^{n_d}\] is such that for every ${A}\in\K^{n_1}\otimes\dots\otimes\K^{n_d}$ \[\pi_i({A})=(U_i^HU_i)\cdot_i{A}\] for every $i\in\{1,\dots,d\}$. Define $P_{{V}}({A}):=\pi_{1}\dots\pi_{d}({A})$. Then for every ${A}\in\K^{n_1}\otimes\dots\otimes\K^{n_d}$ and for every $\rho$ permutation of $d$ elements, we get that:
	\begin{equation}
	\label{eq3_7}
	\norm{{A}-P_{{V}}({A})}^2=\sum_{i=1}^d\norm{\pi_{\rho_{i-1}}\dots\pi_{\rho_1}({A})\,-\, \pi_{\rho_{i}}\dots\pi_{\rho_1}({A})}^2.
	\end{equation}
\end{proposition}
\begin{corollary}
	\label{cor3_4}
	Under the hypothesis of Proposition \ref{prop3_1}, we have that for every tensor ${A}\in\K^{n_1}\otimes\dots\otimes\K^{n_d}$ 
	\[\norm{{A}-P_{{V}}({A})}^2\le\,\sum_{i=1}^d\norm{\pi^\perp_{i}({A})}^2.
	\]
\end{corollary}
Usually, in applications, the multilinear rank of the given tensor is not known. 
Consequently, working on the exact error is not convenient. Therefore, on the basis of Proposition \ref{prop3_1} and Corollary \ref{cor3_4},
N. Vannieuwenhoven K. Meerbergen and R. Vandebril developed two new strategies to approximate tensors, cf.~\cite{mate13}, when the multilinear rank is not known. The first is based on the approximation of the upper bound of the error, stated in Corollary \ref{cor3_4}. As a matter of fact reducing the upper bound implies 
reducing the the exact error.

The \emph{Truncated Higher Order Singular Value Decomposition} (T-HOSVD) has the SVD as key concept. Fixed ${A}\in\K^{n_1}\otimes\dots\otimes\K^{n_d}$,  consider the $i$-th term of the upper bound summation, i.e.
\[0\le \norm{\pi_i^\perp({A})}^2 = \norm[2]{{A}-(U_iU_i^H)_{\cdot_i}{A}}^2.\]
By the positivity of each term of the previous expression, minimizing the upper bound means minimizing each term of the upper bound summation, i.e. taking the minimum of the norm of the difference between ${A}$ and its projection in just one direction each time. Because the tensor is approximated only with respect to one direction each time, the minimization problem is mathematically
\[ \arg\min_{\pi_i\text{ projection into }{V}_i}\norm{\pi_i^\perp({A})}^2=\arg\min_{U_i\in\ort(n_i\times r_i)}\norm[2]{{A}_{(i)}-(U_iU_i^H){A}_{(i)}}^2
,\] i.e. looking for for the best approximation at rank $r_i$ of the $i$-th flattening of ${A}$ for every $i\in\{1,\dots,d\}$. However thanks to Schmidt, Mirsky, Eckhart, Young theorem the  problem for matrices has a close solution which is obtained through SVD of the $i$-th flattening of ${A}$, truncated at the $r_i$ singular values for every $i\in\{1,\dots,d\}$.
It is clear that the order of projectors application is not significant for the T-HOSVD. This won't be true anymore for the \emph{Sequentially Truncated High Order Singular Value} (ST-HOSVD).  
The idea of the ST-HOSVD is minimizing each term of the summation of Proposition \ref{prop3_1}. Let ${A}\in\K^{n_1}\otimes\dots\otimes\K^{n_d}$ be a  tensor and let $(r_1,\dots,r_d)$ be the target multilinear rank of the approximation. The first step is looking for the projector which minimises the first error term of Equation \eqref{eq3_7}:
\begin{align*}
\arg\min_{\pi_1}\norm{\pi_1^\perp({A})}^2&=\arg\min_{U_1\in\ort(n_1\times r_1)}\norm{(U_1U_1^H)_{\cdot_1}{A}}^2,
\end{align*}
i.e.
\[\arg\min_{U_1\in\ort(n_1\times r_1)}\norm{(U_1U_1^H){A}_{(1)}}^2.
\]
The last formulation of this first step has a close solution, thanks to Schmidt, Mirsky, Eckhart, Young theorem. We are looking for the best rank $r_1$ approximation of the $1$-st flattening. So we can conclude that the matrix $U_1$ obtained from the SVD of the first flattening of ${A}$ truncated at the $r_1$-th column is such that
\[\hat{U}_1 =\arg\min_{\substack{U_1\in\ort(n_1\times r_1)}}\norm{(U_1U_1^H){A}_{(1)}}^2.
\] 
The core tensor of the first step,
\[{C}^{(1)} = (\hat{U}_1^H,I,\dots,I)\cdot {A}.
\]
Fixed the $\hat{\pi}_1=\hat{U}_1\hat{U}_1^H$, the next step is looking for the projector which minimises the second error term of Equation \eqref{eq3_7}:
\[\arg\min_{\pi_2}\norm{\pi_2^\perp\hat{\pi}_1({A})}^2=\arg\min_{U_2\in\ort(n_2\times r_2)}\norm{(U_2U_2^H)_{\cdot_2}(\hat{U}_1\hat{U}_1^H)_{\cdot_1}{A}}^2.
\]
But thanks to multilinearity, the last equation becomes
\[\arg\min_{U_2\in\ort(n_2\times r_2)}\norm{(\hat{U}_1)_{\cdot_1}(U_2U_2^H)_{\cdot_2}(\hat{U}_1^H)_{\cdot_1}{A}}^2\]i.e.\[\arg\min_{U_2\in\ort(n_2\times r_2)}\norm{(\hat{U}_1)_{\cdot_1}(U_2U_2^H)_{\cdot_2}{C}^{(1)}}^2.
\]
Since $\hat{U}_1$ is fixed, the second step problem becomes
\[\arg\min_{U_2\in\ort(n_2\times r_2)}\norm{(U_2U_2^H)_{\cdot_2}{C}^{(1)}}^2=\arg\min_{U_2\in\ort(n_2\times r_2)}\norm{(U_2U_2^H)_{\cdot_2}{C}^{(1)}_{(2)}}^2
\]
whose solution is the matrix $\hat{U}_2$ obtained through the SVD of the $2$-nd flattening of ${C}^{(1)}$ truncated at the $r_2$-th column. Defined the new core tensor \[{C}^{(2)}=(\hat{U}^H_1,\hat{U}^H_2,I,\dots,I)\cdot {A},\] 
we proceed similarly with the next ones. At the $(d-1)$-th step the $(d-1)$-th core tensor is defined as
\[{C}^{(d-1)}=(\hat{U}^H_1,\hat{U}^H_2,\dots,\hat{U}^H_{d-1},I)\cdot {A}.
\] For the last $d$-th step, we look for the projectors which minimises the last error term of Equation \eqref{eq3_7}:
\[\arg\min_{\pi_d}\norm{\pi_d^\perp\hat{\pi}_{d-1}\dots\hat{\pi}_1({A})}^2,
\] 
which using the multilinearity and the projectors properties becomes
\[\arg\min_{\substack{U_d\in\ort(n_d\times r_d)}}\norm{(U_dU_d^H)_{\cdot_d}(\hat{U}_1\hat{U}_1^H,..,\hat{U}_{d-1}\hat{U}_{d-1}^H)_{\cdot_{1,..,d-1}}{A}}^2
\]
i.e.
\begin{equation*}
\arg\min_{U_d\in\ort(n_d\times r_d)}\norm{(U_dU_d^H){C}^{(d-1)}_{(k)}}^2.
\end{equation*}
The solution of the last problem is the matrix $\hat{U}_d$ obtained through the SVD of the $d$-th flattening of the core tensor ${C}^{(d-1)}$ truncated at the $r_d$-th column.

\medskip

We can now state both the T-HOSVD and the ST-HOSVD algorithms, whose implementations are in the Appendix, Section \ref{app:code:mat_THOSVD} and Section \ref{approx_STHOSVD} respectively.

\smallskip

\begin{algorithm}[H]
	\SetAlgoLined
	\KwIn{a tensor ${A}\in\K^{n_1}\otimes\dots\otimes\K^{n_d}$}
	\KwIn{a target multilinear rank $(r_1,\dots,r_d)$}
	\KwOut{the T-HOSVD basis in matrix form $(\overline{U}_1,\dots,\overline{U}_d)$}
	\KwOut{the T-HOSVD core tensor ${C}\in\K^{n_1}\otimes\dots\otimes\K^{n_d}$}
	\For{$i = 1,\,2,\dots,d$}{
		$\;\,$Compute SVD of ${A}_{(i)}$, i.e. ${A}_{(i)}= U_i\Sigma_iV_i^T$;\\
		Store in $\overline{U}_i$ the first $r_i$ columns of $U_i$
	}
	${C}\gets (\overline{U}^H_1,\dots,\overline{U}^H_d)\cdot {A}$;
	\caption{T-HOSVD}\label{T-HOSVD_algo}
\end{algorithm}

\bigskip

\begin{algorithm}[H]
	\SetAlgoLined
	\KwIn{a tensor ${A}\in\K^{n_1}\otimes\dots\otimes\K^{n_d}$}
	\KwIn{a target multilinear rank $(r_1,\dots,r_d)$}
	\KwOut{the ST-HOSVD basis in matrix form $\hat{U}_1,\dots,\hat{U}_d$}
	\KwOut{the ST-HOSVD core tensor ${C}\in\K^{n_1}\otimes\dots\otimes\K^{n_d}$}
	${C}\gets {A}$;\\
	\For{$i = 1,\,2,\dots,d$}{
		$\;\,$Compute SVD of ${C}_{(i)}$, i.e. ${C}_{(i)}= U_i\Sigma_iV_i^T$;\\
		Store in $\hat{U}_i$ the first $r_i$ columns of $U_i$;\\
		${C}\gets (\hat{U}^H_i)_{\cdot_i}{C}$;		
	}
	\caption{ST-HOSVD}\label{algoSTHOSVD}
\end{algorithm}

\section{Results}\label{sec:res}

In this section we will first describe the data used: RED, NIR and NDVI bands of different territories, provided by NASA, ~\cite{res0,res00}. The first step in our analysis will be computing the biodiversity index over the NASA NDVI imageries. Then we will generate a new NDVI from RED and NIR using Definition \ref{def2_3_1}, since we do not know how NASA generates NDVI from the other two bands. Over these ``relative" NDVI we will compute the biodiversity indexes. Next we will generate 3-order tensors with just RED and NIR bands. Then we will approximate these tensors with T-HOSVD and ST-HOSVD. Lastly from the approximated tensors we will extract RED and NIR imageries to get ``approximated" NDVI. Over these last NDVI we compute the biodiversity indexes. In conclusion we will measure the error made in estimating biodiversity when we use approximated NDVI instead of NASA NDVI or relative NDVI.     
\subsection{Data}
From MOD13C2v006 and MOD13A3v006, both sensed by MODIS, cf.~\cite{mod1}, and available at~\cite{res0,res00}, we select the RED, NIR rasters and the NASA computed NDVI.\\
MOD13C2v006 is a product characterized by $13$ layers, each of which stores an imagery of the Earth with different properties. Each raster is a matrix of $3600$ rows and $7200$ columns. The side of each pixel corresponds to $\SI{5600}{\metre}$, which is the spatial resolution. The imageries are monthly: they are obtained from the daily data through NASA's algorithms. The chosen data from January 2010 until December 2018 are in \texttt{hdf} format.\\
MOD13A3v006 is a similar product with a higher spatial resolution. While each image from the previous dataset covered the entire Earth's surface, each one from this second dataset covers just $1200\times \SI{1200}{\metre\squared}$. We select the $20$ components, called \emph{granules}, to compose an Europe's map. We download in \texttt{GEOTiff} the granules from RED, NIR bands and from the NDVI computed by the NASA, from June 2011 until December 2018. Also in this case they are obtained by the NASA scientists from daily data. The final dimension of most of the rasters is $4800$ rows and $6000$ columns. We do not talk about every used test elements from MOD13A3v006 dataset, since those of December 2016 and of December 2017 do not include all the $20$ granules. Moreover we remark that rasters of December 2012 and December 2015 have the correct dimension, but they have respectively one and two missing areas. Lastly NASA stores the data in $16$-bit signed integers.\\ The next step is creating $3$-order tensors. Taking advantage of the GDAL dependencies for python, we simply convert rasters into matrix, removing the metadata useless for our aims. Then we store NDVI, RED and NIR into $3$-order arrays.
\begin{definition}
	Let $\set{T}_E$ be the set of all the tensor ${A}\in\R^{n_1}\otimes\R^{n_2}\otimes\R^3$ with $n_1=4800$ and $n_2=6000$ such that
	\begin{itemize}
		\item ${A}_{\cdot,\cdot,1}$ is the NDVI raster, 
		\item ${A}_{\cdot,\cdot,2}$ is the RED raster band,
		\item${A}_{\cdot,\cdot,3}$ is the NIR raster band 
	\end{itemize}
	of Europe dataset for every month and for every year. The cardinality of $\set{T}_E$ is $n_E=91$. \\
	Similarly let $\set{T}_W$ be the set of all the tensor ${A}\in\R^{m_1}\otimes\R^{m_2}\otimes\R^3$ with $m_1=3600$ and $m_2=7200$ such that
	\begin{itemize}
		\item ${A}_{\cdot,\cdot,1}$ is the NDVI raster, 
		\item ${A}_{\cdot,\cdot,2}$ is the RED raster band,
		\item${A}_{\cdot,\cdot,3}$ is the NIR raster band 
	\end{itemize}
	of Earth dataset for every month and for every year. The cardinality of $\set{T}_W$ is $n_W=108$.
\end{definition}
\begin{remark}
	Henceforth we will denote with $\R^{\otimes E}$ the tensor space $\R^{n_1}\otimes\R^{n_2}\otimes\R^3$, where $E=(n_1,n_2,3)$, $n_1=4800$ and $n_2=6000$ and with $\R^{\otimes W}$ the tensor space $\R^{m_1}\otimes\R^{m_2}\otimes\R^3$, where $W=(m_1,m_2,3)$, $m_1=3600$ and $m_2=7200$. 
\end{remark}
Since obtaining biodiversity indexes takes long time and needs many resources, we perform all the computation over HPC@UniTrento, the university of Trento cluster. The indexes measured are Rao and Rényi both with window side equal to $11$. We maintain the same window side on Europe's and on Earth's images, since the different spatial resolutions lead to similar rasters dimensions.
\subsection{NASA and relative NDVI}
The first step is computing both Rao and Rényi indexes over the NDVI raster, extracted from the loaded tensor, i.e. over $({A}_{k_h})_{\cdot,\cdot,1}$ for every $k_h\in\{1,\dots,n_h\}$ for every $h\in\{E,W\}$.
\begin{definition}
	Let $\set{R}_{h}$ be the set of Rao index computed over $({A}_{k_h})_{\cdot,\cdot,1}$ for every $k_h\in\{1,\dots,n_h\}$ for every $h\in\{E,W\}$. Similarly let $\set{I}_{h}$ be the set of Rényi index computed over $({A}_{k_h})_{\cdot,\cdot,1}$ for every $k_h\in\{1,\dots,n_h\}$ for every $h\in\{E,W\}$.\\
	We call $R$ \emph{original estiamtes} for every $R\in\set{R}_{E}\cup\set{R}_W\cup\set{I}_E\cup\set{I}_W$.
\end{definition} 
We use the obtained images as comparison term. Then we compute also an NDVI from the RED and NIR raster, using Definition \ref{def2_3_1}, since the algorithm used by NASA for NDVI creation is not public. 
\begin{definition}
	\label{def3_2_1}
	Let $g_E:\R^{\otimes E}\mapsto \M^{n_1\times n_2}(\R)\times \M^{n_1\times n_2}(\R)$ and $g_W:\R^{\otimes W}\mapsto \M^{m_1\times m_2}(\R)\times \M^{m_1\times m_2}(\R)$ be such that
	\[g_h({A})=({A}_{\cdot,\cdot,2},{A}_{\cdot,\cdot,3})
	\]
	for every ${A}\in\set{T}_h$ for every $h\in\{E,W\}$.
	Let $M\in\R$ be a default missing value and let $l:\R\times \R\mapsto \R$ be a function such that
	\begin{equation*}
	l(a,b)=
	\begin{cases}
	\frac{a-b}{a+b}\quad&\text{if }a+b\ne 0\\
	M\quad&\text{if }a+b=0
	\end{cases}
	\end{equation*}
	for every $a,b\in \R$.
	Let $\overline{l}:\M^{p\times q}(\R)\times\M^{p\times q}(\R)\mapsto \M^{p\times q}(\R)$ be such that for every $A,B\in\M^{p\times q}(\R)$ then $\overline{l}(A,B)=C$ such that $C_{i,j}=l(A_{i,j},B_{i,j})$ for every $i\in\{1,\dots,p\}$ for every $j\in\{1,\dots,q\}$.\\ \smallskip
	Let $f_h=\overline{l}\circ g_h$ be the function that associates to each tensor ${A}\in\set{T}_h$ the NDVI raster obtained from $({A}_{k_h})_{\cdot,\cdot,2}$ and $({A}_{k_h})_{\cdot,\cdot,3}$ for every $k_h\in\{1,\dots,n_h\}$ for every $h\in\{E,W\}$. Then \[\tilde{\set{T}}_h = f_h(\set{T}_h)\] for every $h\in\{E,W\}$.
	
	We call elements of $\set{N}_h$ \emph{self-made NDVI images} for every $h\in\{E,W\}$.
\end{definition}
\begin{remark}
	Since rasters have only non negative elements for every NIR and RED band, the second case of function $l$ in the previous definition is verified when both elements of NIR and RED rasters are zero. In that case we assign to NDVI the default missing value, $M=-3000$.
\end{remark}
Consequently we perform again index estimation over $\set{N}_h$ elements for every $h\in\{E,W\}$.
\begin{definition}
	Let $\set{\tilde{R}}_{h}$ be the set of Rao index computed over $A_{k_h}\in\set{N}_h$ for every $k_h\in\{1,\dots,n_h\}$ for every $h\in\{E,W\}$. Similarly let $\set{\tilde{I}}_h$ be the set of Rényi index computed over $A_{k_h}\in\set{N}_h$ for every $k_h\in\{1,\dots,n_h\}$ for every $h\in\{E,W\}$.
	
	We call $R$ \emph{relative estiamtes} for every $R\in\set{\tilde{R}}_{E}\cup\set{\tilde{R}}_W\cup\set{\tilde{I}}_E\cup\set{\tilde{I}}_W$.
\end{definition} 
\begin{remark}
	Henceforth we assume that elements of the same set pairs $(\set{R}_h,\set{\tilde{R}}_h)$ and $(\set{I}_h,\set{\tilde{I}}_h)$ are ordered equally for every $h\in\{E,W\}$. 
\end{remark}
Even now, we can present some comparison between these two types of estimates. We compute the error $\norm{A_j-B_j}_F$ for every $A_j\in\set{R}_i$ and $B_j\in\set{\tilde{R}}_i$ for every $j\in\{1,\dots,n_i\}$ for every $i\in\{E,W\}$. Next we also compute the error per pixel dividing the error by the number of pixels. Since we are working with huge matrices, this type of distributed error is useful to understand how big is on average the error made pointwise. 
In the following table we present some statistics, where $e$ is the error vector and $ep$ is the error per pixel vector. Besides for every $v\in\K^n$, we set
\[\min v = \min\{v_1,\dots,v_n\}\quad\text{and}\quad\max v=\max\{v_1,\dots,v_n\}.
\] 
\begin{table}[H]
	\begin{center}
		\label{tab3_1}
		\begin{tabular}{ccccccccccc} 
			\toprule
			\textbf{Dataset}&$\E[e]$&$\E[ep]$&$\V[e]$&$\V[ep]$&$\min ep$&$\max ep$\\
			\midrule
			Europe&$197877.819$&$0.0069$&$409993378560.6464$&$0.0005$&$0.0013$&$0.1777$\\
			\hline
			World&$1731817.3949$&$0.0668$&$687978783275.7339$&$0.001$&$0.0155$&$0.1678$\\
			\bottomrule
		\end{tabular}
		\caption{Rao index statistics for original and relative data.}
	\end{center}
\end{table}
We make on average a $0.6\%$ error per pixel for Europe Rao estimation using self-made NDVI instead of NASA NDVI, while the error is on average of $6\%$ per pixel when Rao is computed over Earth NDVIs. Both the unbiased variance are very small, which means that errors are not very different from the mean.
Notice that also the minimum approximation error is greater for Earth than for Europe data. We suppose that this is linked with the water surface greater in World rasters than in Europe's ones. \\
Similarly we compute the error $\norm{A_j-B_j}_F$ for every $A_j\in\set{I}_i$ and $B_j\in\set{\tilde{I}}_i$ for every $j\in\{1,\dots,n_i\}$ for every $i\in\{E,W\}$.
With the same notation already introduced, we present some statistics for Rényi index.
\begin{table}[H]
	\begin{center}
		\label{tab3_2}
		\begin{tabular}{cccccccccc} 
			\toprule
			\textbf{Dataset}&$\E[e]$&$\E[ep]$&$\V[e]$&$\V[ep]$&$\min ep$&$\max ep$\\
			\midrule
			Europe&$132517.8038$&$0.0046$&$420678175017.4135$&$0.0005$&$0.0003$&$0.1777$\\
			\hline
			World&$782043.4085$&$0.0302$&$148701528527.3207$&$0.0002$&$0.0112$&$0.0688$
			\\
			\bottomrule
		\end{tabular}
		\caption{Rényi index statistics for original and relative data.}
	\end{center}
\end{table}
In this case the mean error per pixel for both the dataset is smaller than the previous mean. One possible explication could be that Rényi index takes into account only frequencies while Rao index considers both values and their frequencies. However in this case variance in World error is smaller than in Europe, while in the Rao case there is the opposite situation. Moreover we observe that the minimum and the maximum approximation error in the World dataset for both the indexes is realised by the same element, i.e. April 2018 and May 2014 respectively. This consideration holds also for Europe dataset: the minimum error comes from April 2018 and the maximum from December 2012.

\subsection{Approximated NDVI}
Before applying the approximation tensors codes, described in  the Appendix, Sections  \ref{app:code:mat_THOSVD}  and \ref{approx_STHOSVD}, we have to highlight one limit of the python function \texttt{svds}. 
It takes as additional parameter \texttt{k}, which is the rank of the wanted approximation and which has to be strictly lower than both the dimensions of the given matrix. So if we had passed just a $3$-order tensor such that $n_3 = 2$, for the third flattening we would have fixed \texttt{k} equal to $1$, getting a vector: this is a too low order tensor for our aims. Therefore we decide to increase $n_3$ up to $3$, adding another matrix to our tensor: in the first case we take twice RED band raster and once NIR one, in the second case we take twice NIR band and once RED one.
\begin{definition}
	Let $g_{R,h}:R^{\otimes h}\mapsto\R^{\otimes h}$ be the function that associate the tensor ${B}$ such that
	\[{B}_{\cdot,\cdot,1}={B}_{\cdot,\cdot,2}={A}_{\cdot,\cdot,2}\quad\text{and}\quad{B}_{\cdot,\cdot,3}={A}_{\cdot,\cdot,3},\] to each tensor ${A}\in\set{T}_h$ for every $h\in\{E,W\}$. Then \[\set{T}_{R,h} = g_{R,h}(\set{T}_h)\] for every $h\in\{E,W\}$. 
	
	Similarly let $g_{N,h}:R^{\otimes h}\mapsto\R^{\otimes h}$ be the function that associates the tensor ${B}$ such that
	\[{B}_{\cdot,\cdot,1}={B}_{\cdot,\cdot,3}={A}_{\cdot,\cdot,3}\quad\text{and}\quad{B}_{\cdot,\cdot,2}={A}_{\cdot,\cdot,2},\]to each tensor ${A}\in\set{T}_h$ for every $h\in\{E,W\}$. Then \[\set{T}_{N,h} = g_{N,h}(\set{T}_h)\] for every $h\in\{E,W\}$,
\end{definition}
To compute T-HOSVD and ST-HOSVD, we fix five multilinear target ranks.
\begin{definition}
	Let $\set{R}=\{10,50,100,500,1000\}$ be a set with the given order fixed, then the \emph{target multilinear rank} we choose are
	\[r_j=(i_j,i_j,2)
	\]
	for every $i_j\in\set{R}$ for every $j\in\{1,\dots,5\}$.
	Let $\set{T}_{k,h,j}$ be the set of T-HOSVD approximation at multilinear rank $r_j$ of tensors from the set $\set{T}_{k,h}$ for every $h\in\{E,W\}$, for every $k\in\{N,R\}$ and for every $j\in\{1,\dots,5\}$. Similarly let $\set{S}_{k,h,j}$ be the set of ST-HOSVD approximation.
\end{definition}
Before presenting results related to indexes computation, we have some data about storage memory use. Since it depends on the core tensor and on the projectors dimensions, which are equal for T-HOSVD and ST-HOSVD, we report only a table for each dataset. For each tensor ${A}\in\set{T}_{k,W}$ the ratio between the memory used for storing the core tensor and the projectors over the memory used for storing ${A}$ is the same, for every $k\in\{N,R\}$. For tensors of $\set{T}_{k,E}$ it holds the same, except for those elements composed by a lower number of granules, for every $k\in\{N,R\}$. Since they are $2$ over $91$, we neglect them and in the table we present the ratios of memory usage for each rank approximation. We call these ratios \emph{absolute compression ratios}, because they have as denominator the memory used to store two time RED band and once NIR band, or vice-versa. In the following table they are present as \emph{Abs. R}, when RED band raster is repeated and as \emph{Abs. N}, in the other case. Beside we list also a \emph{relative compression ratio}, where the denominator is the amount of memory necessary to store once RED and once NIR band. In the table it is \emph{Rel}. Moreover here and all along the paper for simplicity we write as rank only the significant components of the multilinear rank, i.e. $i_j$ for every $i_j \in\set{R}$ for every $j\in\{1,\dots,5\}$.
\begin{table}[H]
	\centering 
	\begin{tabular}{cccccccc} %
		\toprule
		\multirow{2}{*}{
			\parbox[c]{.2\linewidth}{\centering \textbf{Rank}}}
		& \multicolumn{3}{c}{\textbf{Europe}} &&
		\multicolumn{3}{c}{\textbf{Earth}} \\ 
		\cmidrule{2-4} \cmidrule{6-8}
		
		& {\centering Rel} & {Abs. R}& {Abs. N} && {\centering Rel} & {Abs. R}& {Abs. N} \\
		\midrule
		$10$&$0.0019$&$0.0013$&$0.0013$&&$0.0021$&$0.0014$&$0.0014$ \\
		\hline
		$50$&$0.0095$&$0.0063$&$0.0063$&&$0.0105$&$0.007$&$0.007$\\
		\hline
		$100$&$0.0191$&$0.0127$&$0.0127$&&$0.0212$&$0.0141$&$0.0141$\\
		\hline
		$500$&$0.1024$&$0.0683$&$0.0683$&&$0.1138$&$0.0759$&$0.0759$\\
		\hline
		$1000$&$0.2222$&$0.1481$&$0.1481$&&$0.2469$&$0.1646$&$0.1646$\\
		\bottomrule
	\end{tabular}
	\caption{Rate of compression.} 
	\label{tab3_3}
\end{table}
Remark that even with the greater component wise target multilinear rank, we need just a small percentage of memory with respect the one used for storing the entire tensor. In addiction to this, even the relative ratio at the highest multilinear rank present a significant saving in memory use.\\ 
In order to not get lost during the presentation with the numerous indexes used, we will give a general idea. After having generated new tensors and having approximated them, we  compute new NDVIs through function $f_h$ of Definition \ref{def3_2_1} applied on $\set{T}_{k,h,j}\cup\set{S}_{k,h,j}$ for every $h\in\{E,W\}$, for every $k\in\{N,R\}$ and for every $j\in\{1,\dots,5\}$.  We compute biodiversity index over them. Lastly we measure the difference in estimating biodiversity from approximated NDVI and NASA or self-made NDVI.
\begin{definition}
	Let $\set{N}_{k,h,T,j}=f_h(\set{T}_{k,h,R,j})$ and let $\set{N}_{k,h,S,j}=f_h(\set{S}_{k,h,R,j})$ every $j\in\{1,\dots,5\}$ and for every $h\in\{E,W\}$. We call elements of $\set{N}_{k,h,T,j}\cup\set{N}_{k,h,S,j}$ \emph{approximated NDVIs}.
\end{definition}
\subsubsection{Rényi index}
Next step is computing Rényi index over approximated NDVIs.
\normalcolor
\begin{definition}
	Let $\set{I}_{R,h,k,j}$ be the set of Rényi index computed over elements of $\set{N}_{R,h,t,j}$ for every $h\in\{E,W\}$, for every $t\in\{T,S\}$ and for every $j\in\{1,\dots,5\}$, and  call them $i_j$-\emph{approximated estimates} for every $i_j\in\set{R}$ and for every $j\in\{1,\dots,5\}$.
\end{definition}
\begin{remark}
	Notice that we have $4$ indexes for approximated estimates set: 
	\begin{description}
		\item[$\quad$\nth{1} index] indicates the repeated matrix in the starting tensor R for RED and N for NIR;
		\item[$\quad$\nth{2} index] the belonging dataset, E for Europe and W for Earth;
		\item[$\quad$\nth{3} index]	the approximation algorithm, T for T-HOSVD and S for ST-HOSVD;
		\item[$\quad$\nth{4} index] is associated with the target multilinear rank. 
	\end{description}
\end{remark}
The last step is computing the error with respect to the original estimates, i.e.
\[\norm[2]{A_k - C_{k,j}}
\]
for every $A_k\in\set{I}_h$ and for every $C_{k,j}\in\set{I}_{R,h,T,j}\cup\set{I}_{R,h,S,j}$, for every $k\in\{1,\dots,n_h\}$, for every $j\in\{1,\dots,5\}$ and for every $h\in\{E,W\}$.
Moreover we compute the error with respect to relative estimates, i.e.
\[\norm[2]{B_k - C_{k,j}}
\]
for every $B_k\in\tilde{\set{I}}_h$ and for every $C_{k,j}\in\set{I}_{R,h,T,j}\cup\set{I}_{R,h,S,j}$, for every $k\in\{1,\dots,n_h\}$, for every $j\in\{1,\dots,5\}$ and for every $h\in\{E,W\}$.

Now we present some statistics for the errors per pixel with respect to original estimates, stored in vector $epO$ and relative estimates, in vector $epR$ for Europe dataset, for both the decomposition techniques for each target multilinear rank. In the following table we report as rank only the significant component of the multilinear rank, for simplicity.
\begin{remark}
	To simplify the discussion henceforth and all along the article the $i_j$-\emph{original error} will be the error between original estimate and $i_j$-approximated estimate, while the $i_j$-\emph{relative error} will be the error between relative estimate and $i_j$-approximated estimate for every $i_j\in\set{R}$. 
\end{remark}
\begin{table}[ht]
	\hspace*{- 6mm}
	\centering 
	\footnotesize
	\begin{tabular}{ccccccccccccc} %
		\toprule[1pt]
		& \multicolumn{5}{c}{\textbf{T-HOSVD}} &&
		\multicolumn{5}{c}{\textbf{ST-HOSVD}} \\ 
		\cmidrule{2-6} \cmidrule{8-12}
		
		\textbf{Rank}& $	10	$&$	50	$&$	100	$&$	500	$&$	1000$ && $	10	$&$	50	$&$	100	$&$	500	$&$	1000$\\
		\midrule
		$\E[epO]$	&$	0.1576	$&$	0.0914	$&$	0.0801	$&$	0.0807	$&$	0.0766	$&&$	0.1597	$&$	0.0917	$&$	0.0803	$&$	0.0798	$&$	0.0755	$\\	\hline		
		$\V[epO]$&$	0.0003	$&$	0.0002	$&$	0.0002	$&$	0.0013	$&$	0.0013	$&&$	0.0003	$&$	0.0002	$&$	0.0002	$&$	0.0013	$&$	0.0013	$\\	\hline		
		$\E[epR]$&$	0.1572	$&$	0.0913	$&$	0.0807	$&$	0.0821	$&$	0.079	$&&$0.1593	$&$	0.0918	$&$	0.0811	$&$	0.0815	$&$	0.0779	$\\	\hline		
		$\V[epR]$&$	0.0003	$&$	0.0002	$&$	0.0003	$&$	0.0014	$&$	0.0014	$&&$	0.0003	$&$	0.0002	$&$	0.0003	$&$	0.0014	$&$	0.0014	$\\	\hline				
		$\min epO$&$	0.1164	$&$	0.0641	$&$	0.0537	$&$	0.0466	$&$	0.0383	$&&$	0.1164	$&$	0.0646	$&$	0.0542	$&$	0.0459	$&$	0.0367	$\\	\hline		
		$\min epR$&$	0.1164	$&$	0.0641	$&$	0.0537	$&$	0.0466	$&$	0.0383	$&&$	0.1164	$&$	0.0646	$&$	0.0542	$&$	0.0459	$&$	0.0367	$\\	\hline		
		$\max epO$&$	0.1864	$&$	0.1377	$&$	0.1136	$&$	0.194	$&$	0.1999	$&&$	0.1931	$&$	0.1371	$&$	0.1104	$&$	0.1932	$&$	0.199	$\\	\hline			
		$\max epR$&$	0.1864	$&$	0.147	$&$	0.1618	$&$	0.194	$&$	0.1999	$&&$	0.1931	$&$	0.1479	$&$	0.1679	$&$	0.1932	$&$	0.199	$\\
		
		\bottomrule[1pt]
	\end{tabular}
	\caption{Statistics for Rényi index over $N\in\set{N}_{R,E,t,j}$.} 
	\label{tab_3_REI}
\end{table}
Notice that ST-HOSVD original and relative error per pixel is lower than T-HOSVD original and relative error per pixel, when the first two components of target multilinear rank are greater or equal than $500$. The variance of both errors is quite low, even if it increases in the last three multilinear ranks. Moreover we underline that the minimum relative error and the minimum original error are equal up to the forth decimal digit. They also decrease, when the components of multilinear rank increase. On the other hand the maximum of relative errors and the maximum of original errors do not coincide. Beside, they increase significantly in the forth and fifth approximation. Finally we notice that the average relative error per pixel is frequently slightly greater than the original one. For T-HOSVD this inequality between original and relative error average happens from the third approximation, while for the ST-HOSVD from the second one. The difference seems to grow for increasing multilinear rank components. This could appear a bit strange, since we expect the contrary. However we have to remark that Rényi index takes into account only raster values frequencies, neglecting the values themselves. In addiction from the complete data, we observe that the relative error of elements with missing granules, tensors of December 2012 and December 2015, is more than $3$ times the original error.

We can now list statistics about the Earth dataset. Similarly in vector $epO$ there are the errors per pixel with respect to original Rényi estimates, while in $epR$ with respect to relative estimates for each target multilinear rank. 
\begin{table}[H]
	\hspace*{- 6mm}
	\centering 
	\footnotesize
	\begin{tabular}{ccccccccccccc} %
		\toprule[1pt]
		& \multicolumn{5}{c}{\textbf{T-HOSVD}} &&
		\multicolumn{5}{c}{\textbf{ST-HOSVD}} \\ 
		\cmidrule{2-6} \cmidrule{8-12}
		
		\textbf{Rank}	&$	10	$&$	50	$&$	100	$&$	500	$&$	1000	$&&$	10	$&$	50	$&$	100	$&$	500	$&$	1000	$\\	\hline
		$	\E[epO]	$&$	0.1376	$&$	0.0943	$&$	0.0875	$&$	0.0626	$&$	0.0584	$&&$	0.1383	$&$	0.0946	$&$	0.0876	$&$	0.0625	$&$	0.058	$\\	\hline
		$	\V[epO]	$&$	0.0001	$&$	0.0001	$&$	0.0002	$&$	0.0002	$&$	0.0001	$&&$	0.0001	$&$	0.0001	$&$	0.0002	$&$	0.0002	$&$	0.0001	$\\	\hline
		$	\E[epR]	$&$	0.1358	$&$	0.091	$&$	0.0846	$&$	0.0545	$&$	0.0491	$&&$	0.1365	$&$	0.0912	$&$	0.0847	$&$	0.0545	$&$	0.0485	$\\	\hline
		$	\V[epR]	$&$	0.0001	$&$	0.0001	$&$	0.0002	$&$	0.0002	$&$	0.0	$&&$	0.0001	$&$	0.0001	$&$	0.0002	$&$	0.0002	$&$	0.0	$\\	\hline
		$	\min epO	$&$	0.1156	$&$	0.0758	$&$	0.0668	$&$	0.048	$&$	0.0471	$&&$	0.1156	$&$	0.076	$&$	0.0672	$&$	0.0483	$&$	0.0468	$\\	\hline
		$	\min epR	$&$	0.1155	$&$	0.0748	$&$	0.0655	$&$	0.0421	$&$	0.0414	$&&$	0.1155	$&$	0.075	$&$	0.0658	$&$	0.0421	$&$	0.0411	$\\	\hline
		$	\max epO	$&$	0.1599	$&$	0.1182	$&$	0.1255	$&$	0.0967	$&$	0.0813	$&&$	0.1601	$&$	0.1185	$&$	0.1253	$&$	0.0973	$&$	0.0811	$\\	\hline
		$	\max epR	$&$	0.155	$&$	0.1134	$&$	0.1263	$&$	0.0968	$&$	0.0744	$&&$	0.1551	$&$	0.1138	$&$	0.126	$&$	0.0974	$&$	0.0731	$\\	
		\bottomrule[1pt]
	\end{tabular}
	\caption{Statistics for Rényi index over $N\in\set{N}_{R,W,t,j}$.} 
	\label{tab_3_RWI}
\end{table}
We remark that even in this case on average ST-HOSVD technique leads to lower original and relative errors than T-HOSVD, for the last two and for the last one target multilinear rank respectively. Moreover we observe that relative error mean is slightly lower than original one, as we expected. Minimum and maximum of both errors decrease for increasing multilinear rank components. Certainly the most stunning value is the variance of relative error per pixel at the last approximation. For both T-HOSVD and ST-HOSVD it is lower than $10^{-4}$.\\
We include the images associated to Rényi index in the five approximations for the Europe worst case and the Earth best case, both from ST-HOSVD approximation technique.
\begin{example}
	\label{ex3_2_1}
Looking at the Rényi index computed over Europe NDVI of February 2013 as it is in Figure \ref{fig3_2_1}, 
		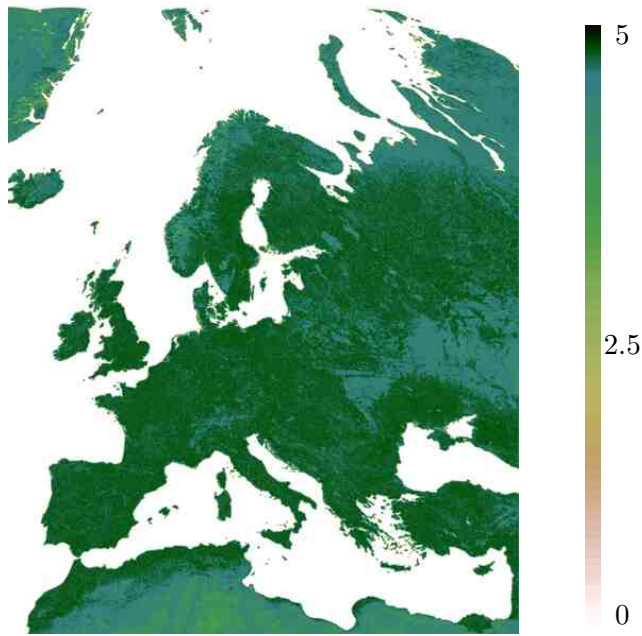
\begin{figure}
			\centering
			\subfloat{	\includegraphics[scale=0.4]{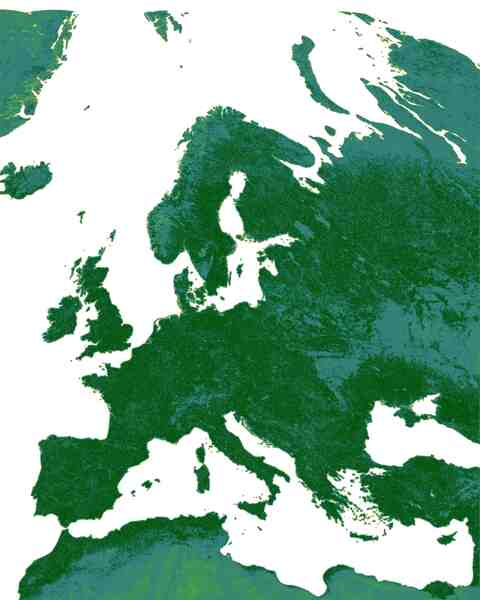}}\qquad
			\subfloat{\begin{tikzpicture}
				\node (img) {\includegraphics[scale=0.47,angle =90, origin=c]{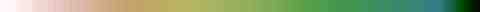}};
				\node [below right,text width=0.5cm,align=center] at (img.north){\normalcolor 5};
				\node [below right,text width=0.5cm,align=center] at (img.center){\normalcolor 2.5};
				\node [above right,text width=0.5cm,align=center] at (img.south){\normalcolor 0};
				\end{tikzpicture}}\\
			\caption{Rényi index computed over NASA NDVI of February 2013.}
			\label{fig3_2_1}
		\end{figure}
	we immediately notice that biodiversity seems to be quite high, near $4.5$ almost everywhere in Europe. However as remarked in~\cite{eco33}, Rényi index tends to overestimate biodiversity. Besides we underline that Rényi index computed over self-made NDVI does not differ much from its computation over NASA NDVI.\\
	Next we have in Figure \ref{fig3_2_2:a} the same index computed over self-made NDVI and the approximated Rényi estimates at different multilinear ranks.
		\begin{figure}
			\centering
			\subfloat[Relative approximation]{\label{fig3_2_2:a}\includegraphics[scale=0.28]{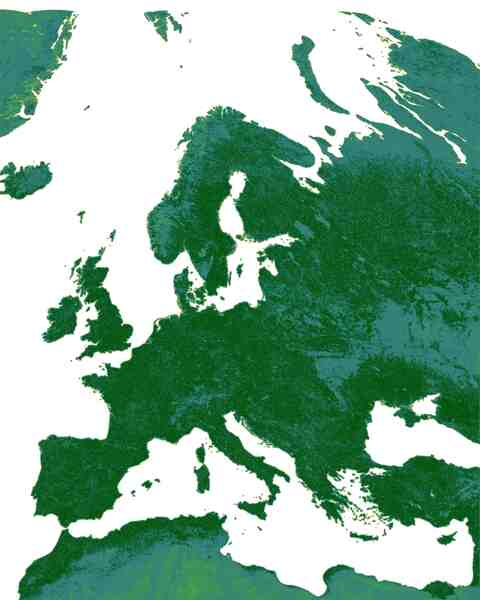}}\qquad
			\subfloat[Component rank $10$ compression]{\label{fig3_2_2:b}\includegraphics[scale=0.28]{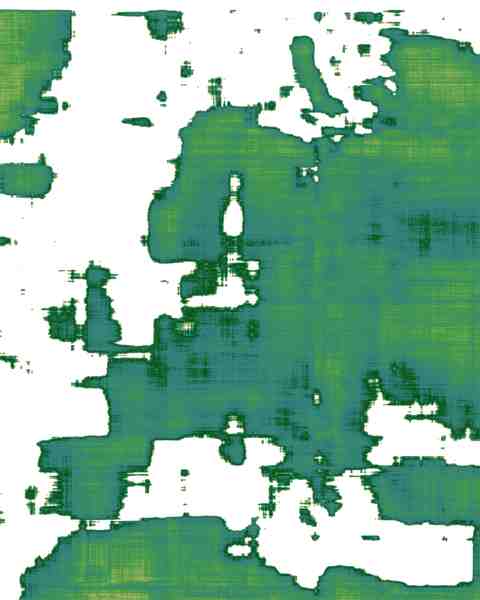}}\qquad
			\subfloat[Component rank $50$ compression]{\label{fig3_2_2:c}\includegraphics[scale=0.28]{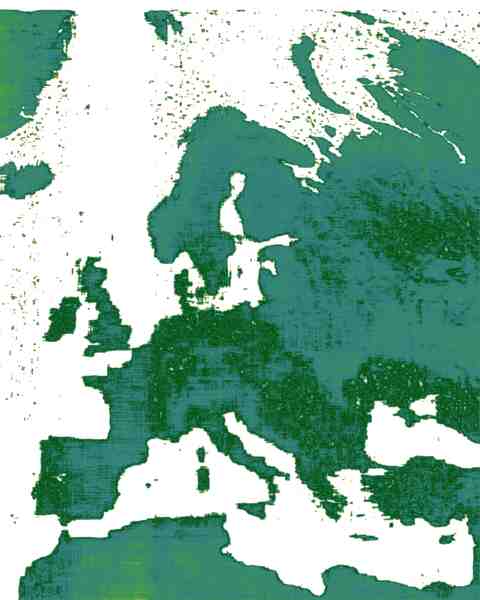}}\\
			\subfloat[Component rank $100$ compression]{\label{fig3_2_2:d}\includegraphics[scale=0.28]{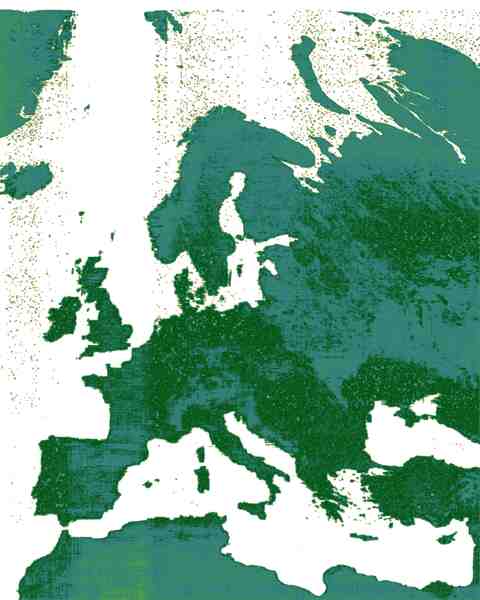}}\qquad
			\subfloat[Component rank $500$ compression]{\label{fig3_2_2:e}\includegraphics[scale=0.28]{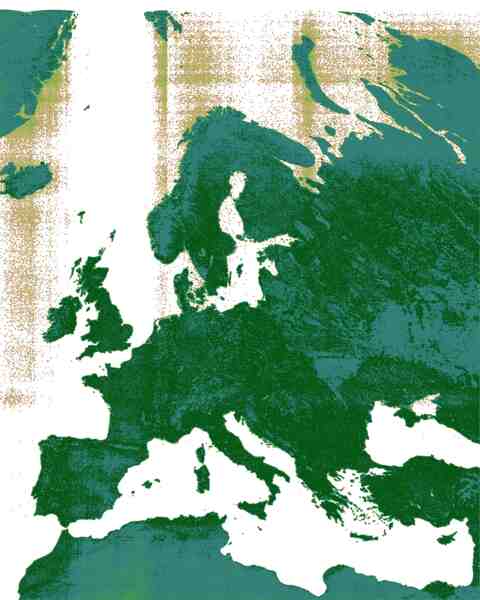}}\qquad
			\subfloat[Component rank $1000$ compression]{\label{fig3_2_2:f}\includegraphics[scale=0.28]{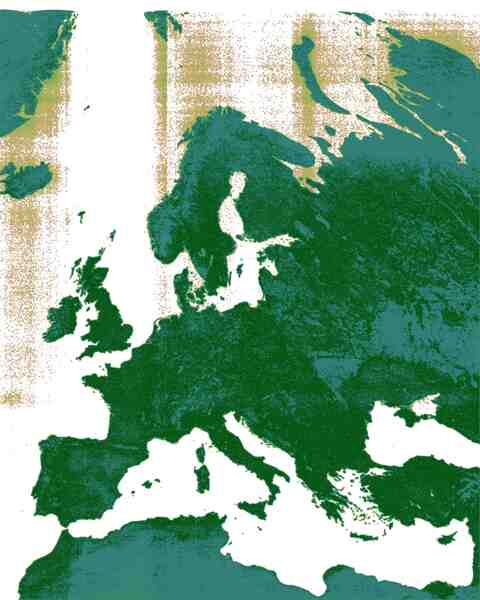}}
			\caption{Approximation of Rényi index for February 2013, from NDVI of $\set{N}_{R,E,S,j}\cup\set{N}_{E}$.}
			\label{fig3_2_2}
		\end{figure} 
	We can notice that when the first two components of the multilinear rank grow, in the Rényi estimation some new noising elements appear, leading to high errors. We believe that this type of phenomenon deserves further analysis. However in the internal land of Europe, the biodiversity estimation is quite close to the relative and original one, for multilinear rank components strictly greater than $100$. 
\end{example}
In the next example we will consider the element of Earth dataset, which realises the minimum error.
\begin{example}
	\label{ex3_2_2}
	Firstly we show in Figure \ref{fig3_2_4} the Rényi estimate over NASA NDVI of October 2017.
		\begin{figure}
		\centering
		\subfloat{\includegraphics[scale=0.45]{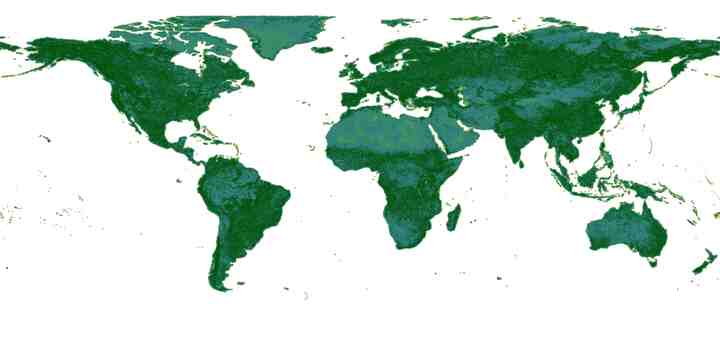}}
			\subfloat{\begin{tikzpicture}
				\node (img) {\includegraphics[scale=0.30,angle =90, origin=c]{PeggiorRenyiERenyiL.jpg}};
				\node [below right,text width=0.5cm,align=center] at (img.north){\normalcolor 5};
				\node [below right,text width=0.5cm,align=center] at (img.center){\normalcolor 2.5};
				\node [above right,text width=0.5cm,align=center] at (img.south){\normalcolor 0};
				\end{tikzpicture}}\\
			\caption{Rényi index computed over NASA NDVI of October 2017.}
			\label{fig3_2_4}
		\end{figure}
	As we said in the Example \ref{ex3_2_1} Rényi index provides quite high biodiversity values. Indeed also in this case there are many Earth areas with a biodiversity value close to $4.5$. Then we present the same index over self-made and approximated NDVI. 
		\begin{figure}
			\centering
			\subfloat[Relative approximation]{\label{fig3_2_5:a}\includegraphics[scale=0.25]{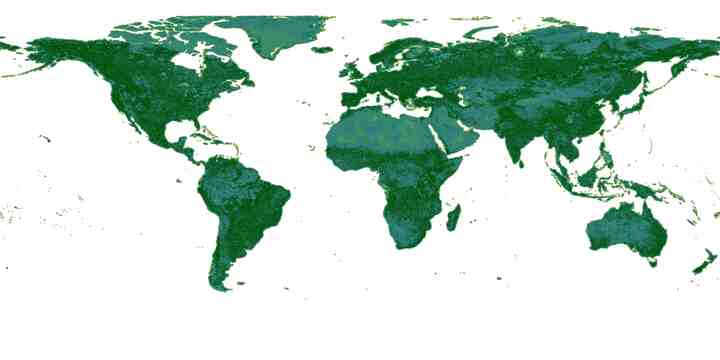}}\qquad
			\subfloat[Component rank $10$ compression]{\label{fig3_2_5:b}\includegraphics[scale=0.25]{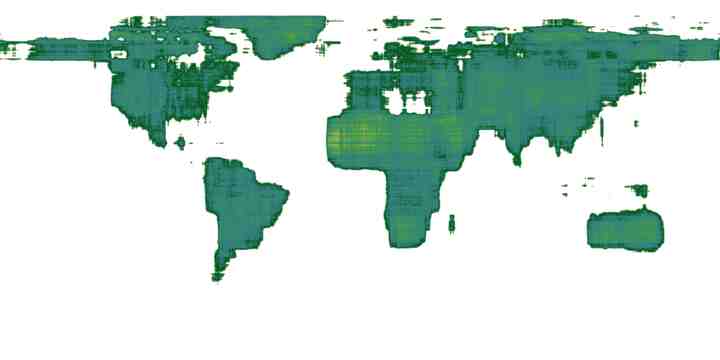}}\\
			\subfloat[Component rank $50$ compression]{\label{fig3_2_5:c}\includegraphics[scale=0.25]{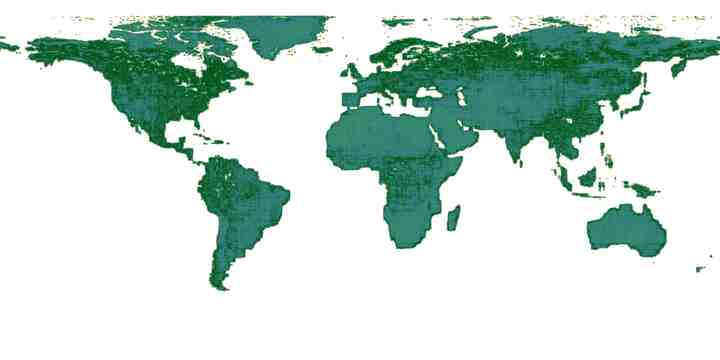}}\quad
			\subfloat[Component rank $100$ compression]{\label{fig3_2_5:d}\includegraphics[scale=0.25]{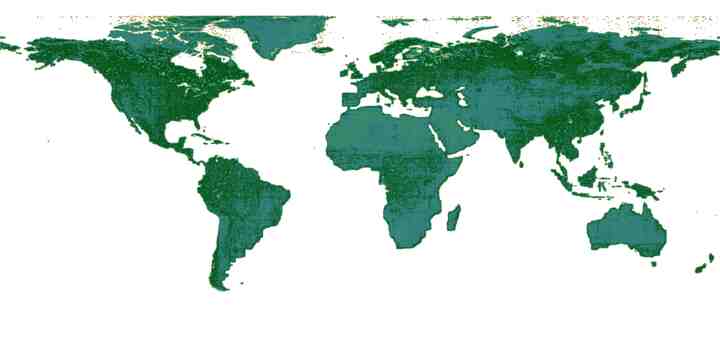}}\\%
			\subfloat[Component rank $500$ compression]{\label{fig3_2_5:e}\includegraphics[scale=0.25]{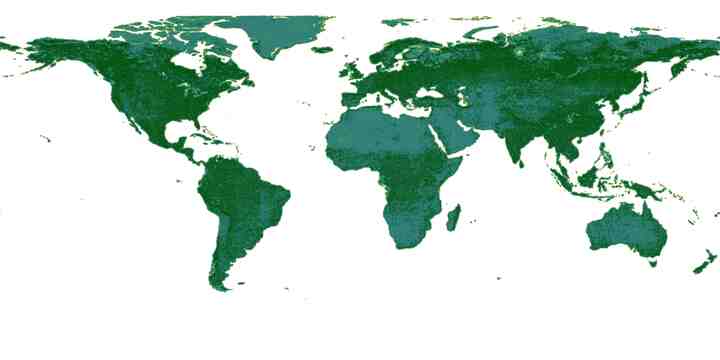}}\qquad%
			\subfloat[Component rank $1000$ compression]{\label{fig3_2_5:f}\includegraphics[scale=0.25]{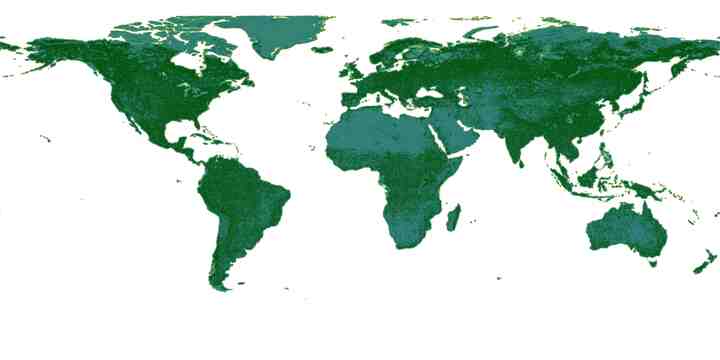}}\\
			\caption{Approximation of Rényi index for October 2017, from NDVI of $\set{N}_{R,W,S,j}\cup\set{N}_{W}$.}
			\label{fig3_2_5}
		\end{figure}
	Even if the small printing dimensions reduce the detail precision, our eyes do not perceive at first glance much difference between the last three index approximations and the index of Figure \ref{fig3_2_4}. However with a closer observation, for example, we notice that islands of the Pacific ocean disappear in first approximation and they partially reappear in the last two images. We may conclude that the original and relative error is in this case linked with these missing territories, but also with the different evaluation of biodiversity in Amazon, for example.
\end{example}
Next step in our discussions is computing Rényi index over approximated NDVI of $\set{N}_{N,h,t,j}$ for every $h\in\{E,W\}$, for every $t\in\{T,S\}$ and $j\in\{1,\dots,5\}$. 
\begin{definition}
	Let $\set{I}_{N,h,t,j}$ be the set of Rényi index computed over elements of $\set{N}_{N,h,t,j}$ for every $h\in\{E,W\}$, for every $t\in\{T,S\}$ and for every $j\in\{1,\dots,5\}$. We decide to call these $i_j$-\emph{approximated estimates} for every $i_j\in\set{R}$ and for every $j\in\{1,\dots,5\}$.
\end{definition}

As before, we compute the error with respect to the original estimates, i.e.
\[\norm[2]{A_k - C_{k,j}}
\]
for every $A_k\in\set{I}_h$ and for every $C_{k,j}\in\set{I}_{N,h,T,j}\cup\set{I}_{N,h,S,j}$, for every $k\in\{1,\dots,n_h\}$, for every $j\in\{1,\dots,5\}$ and for every $h\in\{E,W\}$.
Moreover we compute the error with respect to relative estimates, i.e.
\[\norm[2]{B_k - C_{k,j}}
\]
for every $B_k\in\tilde{\set{I}}_h$ and for every $C_{k,j}\in\set{I}_{N,h,T,j}\cup\set{I}_{N,h,S,j}$, for every $k\in\{1,\dots,n_h\}$, for every $j\in\{1,\dots,5\}$ and for every $h\in\{E,W\}$.\\
Lastly we report some statistics about $i_j$-original and $i_j$-relative errors per pixel, stored in vector $epO$ and $epR$ for each $i_j\in\set{R}$. Firstly a table for Europe related data.
\begin{table}[H]
	\hspace*{- 6mm}
	\centering 
	\footnotesize
	\begin{tabular}{ccccccccccccc} %
		\toprule[1pt]
		& \multicolumn{5}{c}{\textbf{T-HOSVD}} &&
		\multicolumn{5}{c}{\textbf{ST-HOSVD}} \\ 
		\cmidrule{2-6} \cmidrule{8-12}
		\textbf{Rank}	&$	10	$&$	50	$&$	100	$&$	500	$&$	1000	$&&$	10	$&$	50	$&$	100	$&$	500	$&$	1000	$\\	\hline
		$	\E[epO]	$&$	0.1569	$&$	0.09	$&$	0.0782	$&$	0.0794	$&$	0.076	$&&$	0.1586	$&$	0.0903	$&$	0.0783	$&$	0.0787	$&$	0.0751	$\\	\hline
		$	\V[epO]	$&$	0.0003	$&$	0.0002	$&$	0.0002	$&$	0.0013	$&$	0.0013	$&&$	0.0003	$&$	0.0002	$&$	0.0002	$&$	0.0013	$&$	0.0013	$\\	\hline
		$	\E[epR]	$&$	0.1563	$&$	0.0894	$&$	0.0782	$&$	0.0806	$&$	0.0783	$&&$	0.158	$&$	0.0897	$&$	0.0785	$&$	0.08	$&$	0.0775	$\\	\hline
		$	\V[epR]	$&$	0.0003	$&$	0.0002	$&$	0.0002	$&$	0.0014	$&$	0.0015	$&&$	0.0004	$&$	0.0002	$&$	0.0002	$&$	0.0015	$&$	0.0014	$\\	\hline
		$	\min epO	$&$	0.1128	$&$	0.0623	$&$	0.0518	$&$	0.0449	$&$	0.0436	$&&$	0.1143	$&$	0.0625	$&$	0.0521	$&$	0.0438	$&$	0.0428	$\\	\hline
		$	\min epR	$&$	0.1128	$&$	0.0623	$&$	0.0518	$&$	0.0449	$&$	0.0436	$&&$	0.1143	$&$	0.0625	$&$	0.0521	$&$	0.0438	$&$	0.0428	$\\	\hline
		$	\max epO	$&$	0.1827	$&$	0.1462	$&$	0.1275	$&$	0.1936	$&$	0.1991	$&&$	0.1855	$&$	0.1457	$&$	0.1222	$&$	0.1932	$&$	0.1988	$\\	\hline
		$	\max epR	$&$	0.1827	$&$	0.1363	$&$	0.1484	$&$	0.1936	$&$	0.1991	$&&$	0.1855	$&$	0.1373	$&$	0.1528	$&$	0.1932	$&$	0.1988	$\\	
		\bottomrule[1pt]
	\end{tabular}
	\caption{Statistics for Rényi index over $N\in\set{N}_{N,E,t,j}$.} 
	\label{tab_3_NEI}
\end{table}
Almost every consideration for statistics of approximated estimates of $\set{I}_{R,E,t,j}$ for every $t\in\{T,S\}$ and for every $j\in\{1,\dots,5\}$ holds also in this case. However we can notice that on average the $100$ relative and original error are lower that $500$ one, but this is not true anymore for $1000$ relative and original. Moreover when the first two multilinear rank components are strictly smaller that $500$, T-HOSVD relative and original errors are lower than ST-HOSVD ones. For the statistics over $\set{I}_{R,E,t,j}$ elements, this consideration holds only for relative error at the third multilinear rank. The remarks about not full granules elements are true also in this case. We underline that at each multilinear rank both the original and the relative errors on average are smaller in this second case, i.e. applying the procedure to tensors where the NIR band raster is repeated. \\
Lastly some statistical aspects about Earth approximated estimates.
As previously, we list a table with mean, variance, min and max for $i_j$-original and $i_j$-relative errors per pixel, stored respectively in vector $epO$ and $epR$ for each $i_j\in\set{R}$.
\begin{table}[H]
	\hspace*{- 6mm}
	\centering 
	\footnotesize
	\begin{tabular}{ccccccccccccc} %
		\toprule[1pt]
		& \multicolumn{5}{c}{\textbf{T-HOSVD}} &&
		\multicolumn{5}{c}{\textbf{ST-HOSVD}} \\ 
		\cmidrule{2-6} \cmidrule{8-12}
		\textbf{Rank}	&$	10	$&$	50	$&$	100	$&$	500	$&$	1000	$&&$	10	$&$	50	$&$	100	$&$	500	$&$	1000	$\\	\hline
		$	\E[epO]	$&$	0.1351	$&$	0.0915	$&$	0.084	$&$	0.0601	$&$	0.0556	$&&$	0.1355	$&$	0.0917	$&$	0.0842	$&$	0.0601	$&$	0.0553	$\\	\hline
		$	\V[epO]	$&$	0.0001	$&$	0.0001	$&$	0.0002	$&$	0.0002	$&$	0.0001	$&&$	0.0001	$&$	0.0001	$&$	0.0002	$&$	0.0002	$&$	0.0001	$\\	\hline
		$	\E[epR]	$&$	0.1334	$&$	0.0879	$&$	0.0807	$&$	0.052	$&$	0.046	$&&$	0.1338	$&$	0.0882	$&$	0.0809	$&$	0.052	$&$	0.0455	$\\	\hline
		$	\V[epR]	$&$	0.0001	$&$	0.0001	$&$	0.0002	$&$	0.0002	$&$	0.0001	$&&$	0.0001	$&$	0.0001	$&$	0.0002	$&$	0.0002	$&$	0.0	$\\	\hline
		$	\min epO	$&$	0.1119	$&$	0.0727	$&$	0.0638	$&$	0.0443	$&$	0.0424	$&&$	0.112	$&$	0.0728	$&$	0.0641	$&$	0.0445	$&$	0.0422	$\\	\hline
		$	\min epR	$&$	0.1114	$&$	0.0715	$&$	0.0623	$&$	0.0385	$&$	0.0382	$&&$	0.1115	$&$	0.0716	$&$	0.0628	$&$	0.0387	$&$	0.0379	$\\	\hline
		$	\max epO	$&$	0.1564	$&$	0.1154	$&$	0.121	$&$	0.0963	$&$	0.0792	$&&$	0.1574	$&$	0.116	$&$	0.1209	$&$	0.0964	$&$	0.0791	$\\	\hline
		$	\max epR	$&$	0.1536	$&$	0.1098	$&$	0.1218	$&$	0.0965	$&$	0.0731	$&&$	0.1536	$&$	0.1099	$&$	0.1217	$&$	0.0966	$&$	0.0719	$\\	
		\bottomrule[1pt]
	\end{tabular}
	\caption{Statistics for Rényi index over $N\in\set{N}_{N,W,t,j}$.}
	\label{tab_3_NWI}
\end{table} 
Also in this case the considerations presented for $\set{I}_{R,W,t,j}$ error statistics hold. Notice that again variance is lower than $10^{-4}$ for $1000$-relative error of ST-HOSVD. Moreover also in this case ST-HOSVD is convenient only when the first two multilinear rank components are greater than $500$. Lastly original and relative error on average are again smaller in this second case, i.e. when we apply our method to tensors where NIR band is repeated.\\
We will present briefly the Rényi index image over Europe NDVI of February 2013 and over Earth NDVI of October 2017, obtained starting from the correspondent tensors of $\set{T}_{N,h}$ for $h\in\{E,W\}$.
\begin{example}
	In Example \ref{ex3_2_1} we presented the Rényi index image, which realises the highest original and relative error, starting from RED repeated band. Here we have the image of Rényi index for the same element, with approximation obtained from NIR band repeated. We remark that also starting from twice NIR and once RED band tensor element of February 2013 realises the highest original and relative error. 
	\begin{figure}
		\centering
		\subfloat[Component rank $10$ compression]{\label{fig3_2_3:b}\includegraphics[scale=0.28]{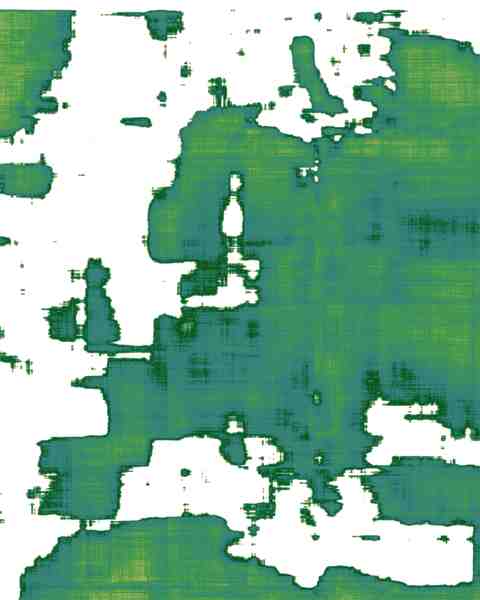}}\qquad
		\subfloat[Component rank $50$ compression]{\label{fig3_2_3:c}\includegraphics[scale=0.28]{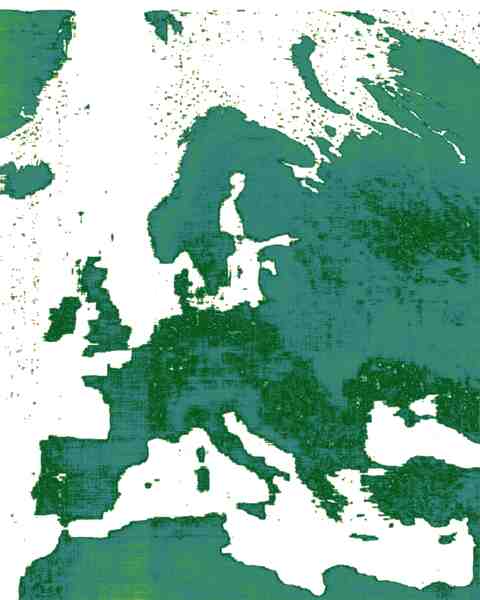}}\qquad
		\subfloat[Component rank $100$ compression]{\label{fig3_2_3:d}\includegraphics[scale=0.28]{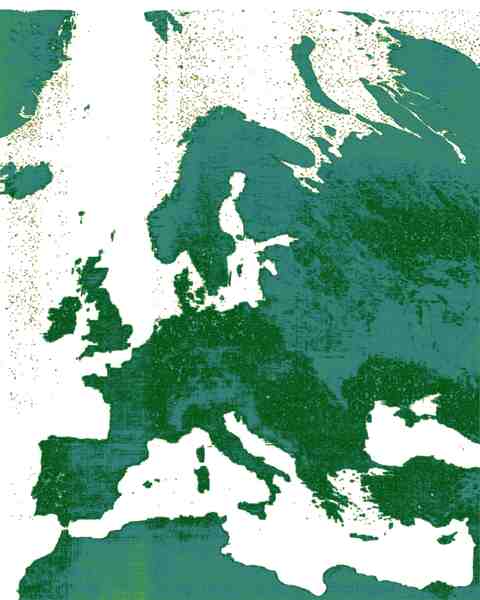}}\\%
		\subfloat[Component rank $500$ compression]{\label{fig3_2_3:e}\includegraphics[scale=0.28]{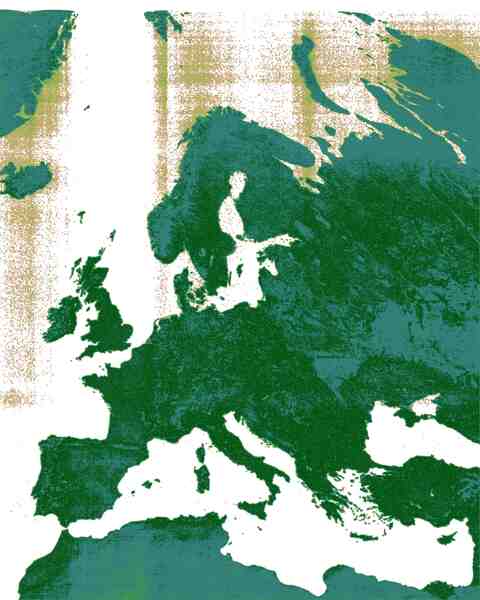}}\qquad%
		\subfloat[Component rank $1000$ compression]{\label{fig3_2_3:f}\includegraphics[scale=0.28]{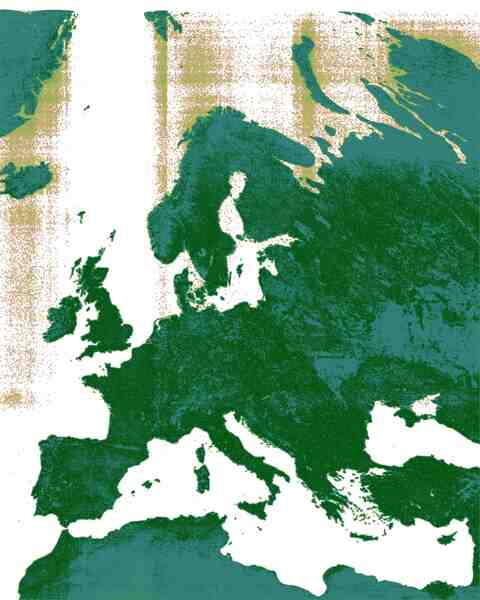}}\qquad
		\subfloat{\begin{tikzpicture}
			\node (img) {\includegraphics[scale=0.32,angle =90, origin=c]{PeggiorRenyiERenyiL.jpg}};
			\node [below right,text width=0.5cm,align=center] at (img.north){\normalcolor 5};
			\node [below right,text width=0.5cm,align=center] at (img.center){\normalcolor 2.5};
			\node [above right,text width=0.5cm,align=center] at (img.south){\normalcolor 0};
			\end{tikzpicture}}
		\caption{Approximation of Rényi index for February 2013, from NDVI of $\set{N}_{R,E,S,j}$.}
		\label{fig3_2_3}
	\end{figure} 
	Again we observe an increasing presence of noise in the north Europe area for growing multilinear rank components. 
\end{example} 
\begin{example}
	Similarly we display approximation of Rényi index for Earth element of October 2017, obtained from tensors where is repeated twice the NIR band.
	\begin{figure}
		\centering
		\subfloat[Component rank $10$ compression]{\label{fig3_2_6:b}\includegraphics[scale=0.25]{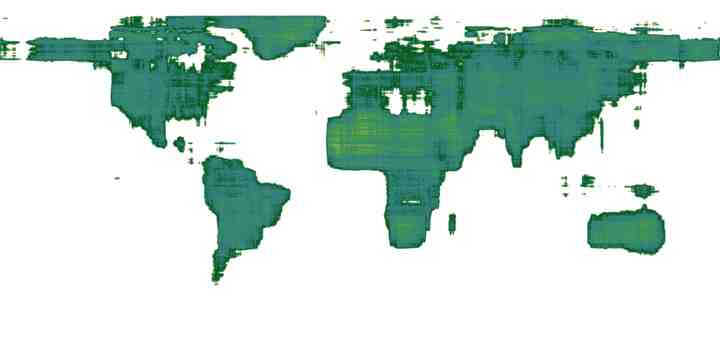}}\quad
		\subfloat[Component rank $50$ compression]{\label{fig3_2_6:c}\includegraphics[scale=0.25]{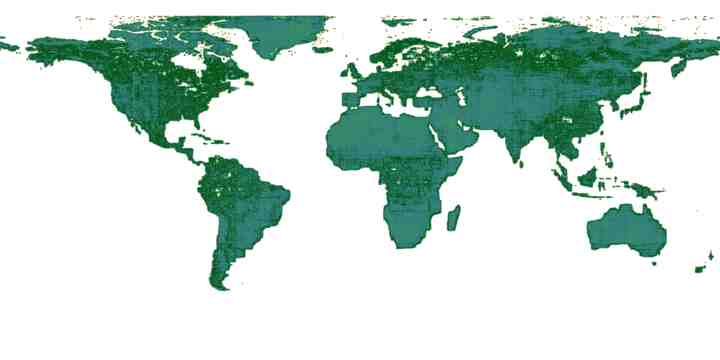}}\\
		\subfloat[Component rank $100$ compression]{\label{fig3_2_6:d}\includegraphics[scale=0.25]{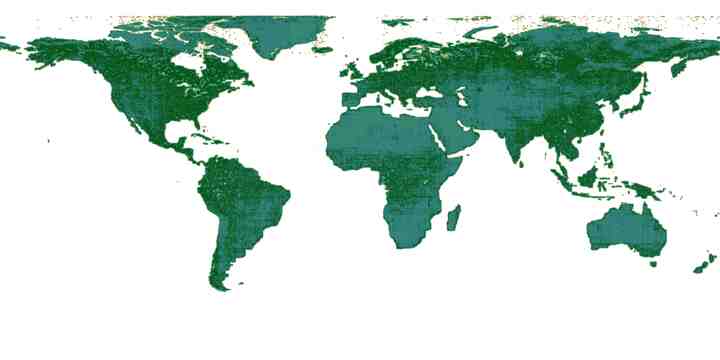}}\qquad%
		\subfloat[Component rank $500$ compression]{\label{fig3_2_6:e}\includegraphics[scale=0.25]{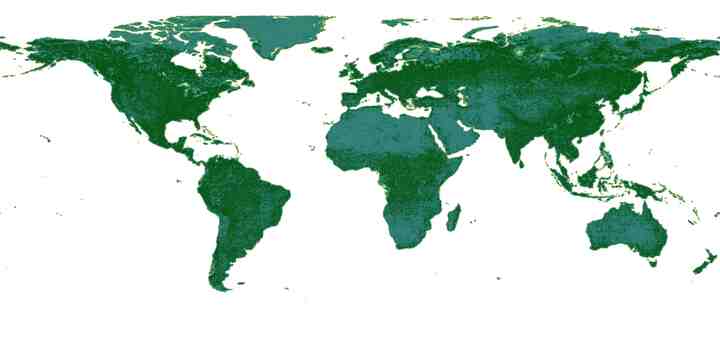}}\\%
		\subfloat[Component rank $1000$ compression]{\label{fig3_2_6:f}\includegraphics[scale=0.45]{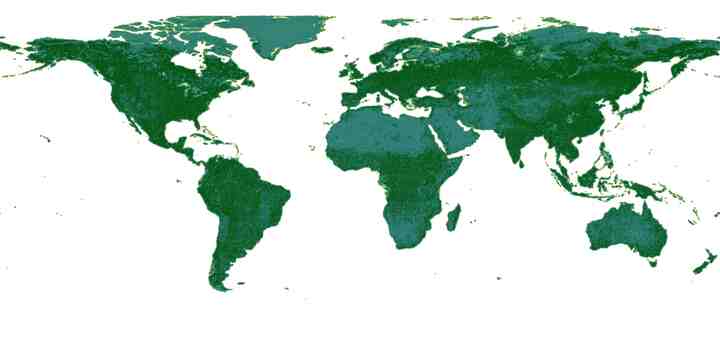}}\qquad
			\subfloat{\begin{tikzpicture}
				\node (img) {\includegraphics[scale=0.20,angle =90, origin=c]{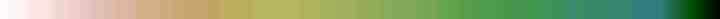}};
				\node [below right,text width=0.5cm,align=center] at (img.north){\normalcolor 5};
				\node [below right,text width=0.5cm,align=center] at (img.center){\normalcolor 2.5};
				\node [above right,text width=0.5cm,align=center] at (img.south){\normalcolor 0};
				\end{tikzpicture}}
			\caption{Approximation of Rényi index for October 2017, from NDVI of $\set{N}_{N,W,S,j}$.}
			\label{fig3_2_6}
		\end{figure}
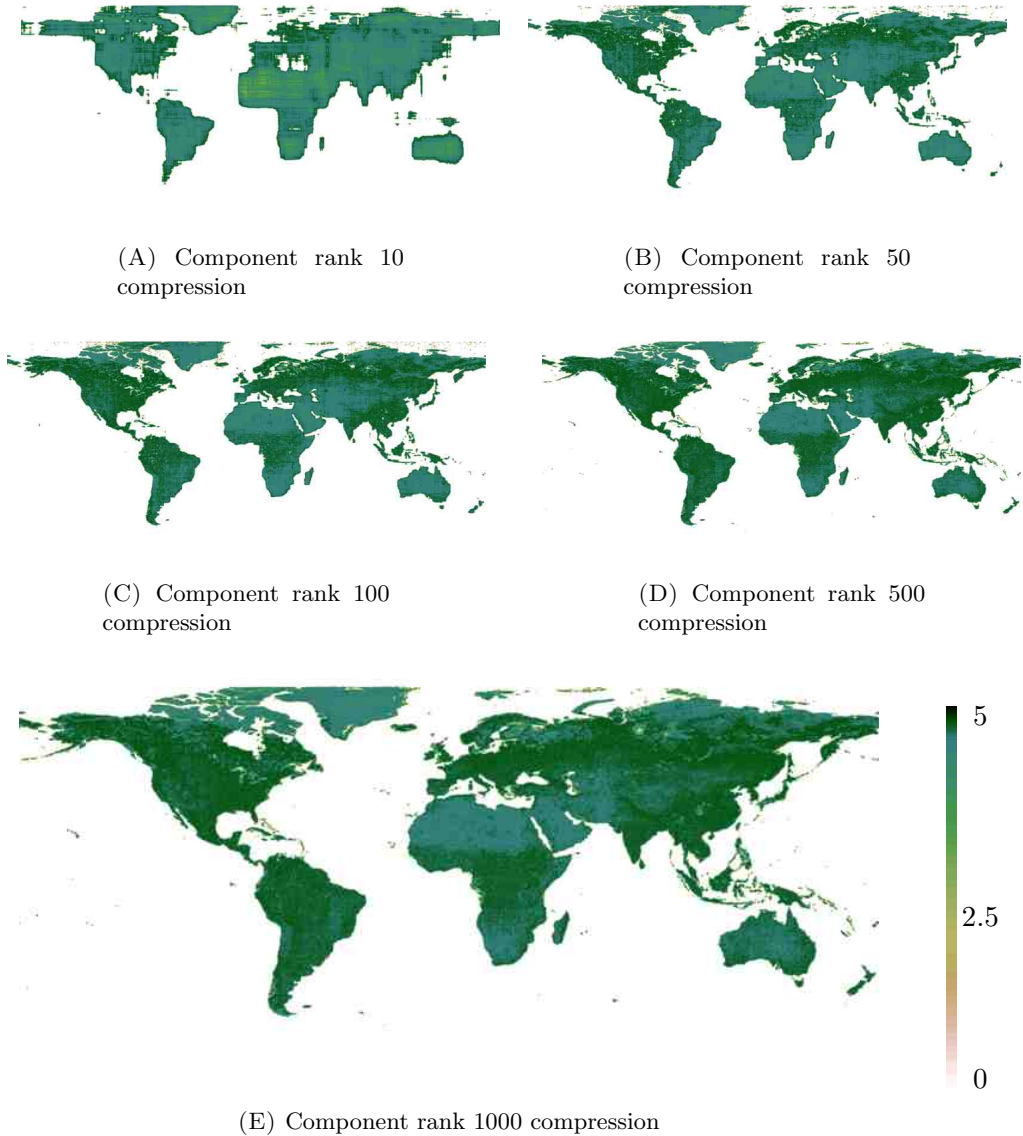
	As in Example \ref{ex3_2_1} we underline that the number of detected territories in the Pacific ocean grows when the first two multilinear rank components grow. Moreover comparing Figure \ref{fig3_2_5:f} and Figure \ref{fig3_2_6:f}, we notice that the amount of Pacific island present is greater in the second one. Indeed this is the element which minimises the original error also in this second proceeding way. Therefore even in this case we can affirm that choosing a starting tensor with repeated NIR band is more convenient that starting with repeated RED band. 
\end{example}
\subsubsection{Rao index} In this second part Rao index is computed over approximated NDVI rasters.
\normalcolor
\begin{definition}
	Let $\set{R}_{R,h,k,j}$ be the set of Rao index computed over elements of $\set{N}_{R,h,t,j}$ for every $h\in\{E,W\}$, for every $t\in\{T,S\}$ and for every $j\in\{1,\dots,5\}$. We call these $i_j$-\emph{approximated estimates} for every $i_j\in\set{R}$ and for every $j\in\{1,\dots,5\}$.
\end{definition}
As we did previously, we compute the error with respect to the original estimates, i.e.
\[\norm[2]{A_k - C_{k,j}}
\]
for every $A_k\in\set{R}_h$ and for every $C_{k,j}\in\set{R}_{R,h,T,j}\cup\set{R}_{R,h,S,j}$, for every $k\in\{1,\dots,n_h\}$, for every $j\in\{1,\dots,5\}$ and for every $h\in\{E,W\}$.
Moreover we compute the error with respect to relative estimates, i.e.
\[\norm[2]{B_k - C_{k,j}}
\]
for every $B_k\in\tilde{\set{R}}_h$ and for every $C_{k,j}\in\set{R}_{R,h,T,j}\cup\set{R}_{R,h,S,j}$, for every $k\in\{1,\dots,n_h\}$, for every $j\in\{1,\dots,5\}$ and for every $h\in\{E,W\}$.\\
To describe our results we report some statistics about $i_j$-original and $i_j$-relative error per pixel, stored in vector $epO$ and $epR$ for each $i_j\in\set{R}$. Firstly we list information for Europe related data.
\begin{table}[h]
	\hspace*{- 6mm}
	\centering 
	\footnotesize
	\begin{tabular}{ccccccccccccc} %
		\toprule[1pt]
		& \multicolumn{5}{c}{\textbf{T-HOSVD}} &&
		\multicolumn{5}{c}{\textbf{ST-HOSVD}} \\ 
		\cmidrule{2-6} \cmidrule{8-12}
		\textbf{Rank}	&$	10	$&$	50	$&$	100	$&$	500	$&$	1000	$&&$	10	$&$	50	$&$	100	$&$	500	$&$	1000	$\\	\hline
		$	\E[epO]	$&$	0.6419	$&$	0.3621	$&$	0.2944	$&$	0.2059	$&$	0.1922	$&&$	0.6328	$&$	0.3604	$&$	0.293	$&$	0.2081	$&$	0.1951	$\\	\hline
		$	\V[epO]	$&$	0.0044	$&$	0.0012	$&$	0.001	$&$	0.0031	$&$	0.003	$&&$	0.0038	$&$	0.0011	$&$	0.001	$&$	0.003	$&$	0.003	$\\	\hline
		$	\E[epR]	$&$	0.6424	$&$	0.3633	$&$	0.2961	$&$	0.2083	$&$	0.1953	$&&$	0.6333	$&$	0.3615	$&$	0.2946	$&$	0.2104	$&$	0.1982	$\\	\hline
		$	\V[epR]	$&$	0.0043	$&$	0.0013	$&$	0.0013	$&$	0.0038	$&$	0.0045	$&&$	0.0037	$&$	0.0013	$&$	0.0013	$&$	0.0038	$&$	0.0045	$\\	\hline
		$	\min epO	$&$	0.4819	$&$	0.274	$&$	0.2194	$&$	0.1306	$&$	0.1128	$&&$	0.4825	$&$	0.2724	$&$	0.2185	$&$	0.1326	$&$	0.1144	$\\	\hline
		$	\min epR	$&$	0.4819	$&$	0.274	$&$	0.2193	$&$	0.1306	$&$	0.1127	$&&$	0.4825	$&$	0.2724	$&$	0.2185	$&$	0.1326	$&$	0.1144	$\\	\hline
		$	\max epO	$&$	0.7559	$&$	0.4081	$&$	0.3591	$&$	0.3858	$&$	0.3894	$&&$	0.7246	$&$	0.4065	$&$	0.3722	$&$	0.3871	$&$	0.3917	$\\	\hline
		$	\max epR	$&$	0.7558	$&$	0.465	$&$	0.4659	$&$	0.4486	$&$	0.5599	$&&$	0.7246	$&$	0.4746	$&$	0.4734	$&$	0.4742	$&$	0.5618	$\\	
		\bottomrule[1pt]
	\end{tabular}
	\caption{Statistics for Rao index over $N\in\set{N}_{R,E,t,j}$.} 
	\label{tab_3_RER}
\end{table}
The most evident aspect is the high mean of both original and relative error made, even when the components of multilinear rank grow. Indeed this average is close to a $20\%$ of error per pixel. If we compare it with the average error per pixel made for Rényi index, we could think that in this case HOSVD is not performant. However we want to remark two critical aspects: firstly Rao index takes into account also the values of NDVI, not only their frequencies. Next we remind that at multilinear rank $r_5$ we are keeping nearly $15\%$ of the total information, as in Table \ref{tab3_3}. Therefore even if results on average are not as good as in Rényi case, we do not exclude the power of this method for Rao index computation. Besides in the best case we have both original and relative error per pixel near to $11\%$, which is appreciable, as we will see. Lastly we remark an interesting and unexpected twist. If in the previous analysis T-HOSVD performed better when multilinear rank components were small with respect to ST-HOSVD, in the present case T-HOSVD provides better result than the other algorithm when the first two multilinear rank components are greater that $500$. Notice also that in this case the relative error is on average greater than the original one. This phenomenon is at least in part linked to the two incomplete elements, December 2012 and 2015. Indeed for these elements the relative error is slightly less than twice the original error.\smallskip

Now we present some statistics for the original and relative errors of Rao estimates for Earth dataset. As before in the following table $epO$ is the vector of the errors per pixel computed with respect to original estimates, while in $epR$ we store the errors per pixel with respect to relative estimates.
\begin{table}[H]
	\hspace*{- 6mm}
	\centering 
	\footnotesize
	\begin{tabular}{ccccccccccccc} %
		\toprule[1pt]
		& \multicolumn{5}{c}{\textbf{T-HOSVD}} &&
		\multicolumn{5}{c}{\textbf{ST-HOSVD}} \\ 
		\cmidrule{2-6} \cmidrule{8-12}
		\textbf{Rank}	&$	10	$&$	50	$&$	100	$&$	500	$&$	1000	$&&$	10	$&$	50	$&$	100	$&$	500	$&$	1000	$\\	\hline
		$	\E[epO]	$&$	0.5442	$&$	0.3476	$&$	0.2949	$&$	0.1845	$&$	0.1735	$&&$	0.5416	$&$	0.3467	$&$	0.2945	$&$	0.1851	$&$	0.1752	$\\	\hline
		$	\V[epO]	$&$	0.0021	$&$	0.0007	$&$	0.0007	$&$	0.0004	$&$	0.0002	$&&$	0.0018	$&$	0.0007	$&$	0.0007	$&$	0.0004	$&$	0.0002	$\\	\hline
		$	\E[epR]	$&$	0.5425	$&$	0.3445	$&$	0.2917	$&$	0.175	$&$	0.1628	$&&$	0.54	$&$	0.3436	$&$	0.2913	$&$	0.1756	$&$	0.1646	$\\	\hline
		$	\V[epR]	$&$	0.0019	$&$	0.0006	$&$	0.0006	$&$	0.0003	$&$	0.0001	$&&$	0.0017	$&$	0.0005	$&$	0.0006	$&$	0.0003	$&$	0.0001	$\\	\hline
		$	\min epO	$&$	0.4614	$&$	0.2943	$&$	0.2472	$&$	0.1551	$&$	0.1523	$&&$	0.4613	$&$	0.294	$&$	0.2475	$&$	0.1555	$&$	0.1539	$\\	\hline
		$	\min epR	$&$	0.4625	$&$	0.2952	$&$	0.2473	$&$	0.1509	$&$	0.1454	$&&$	0.4623	$&$	0.295	$&$	0.2473	$&$	0.1513	$&$	0.1473	$\\	\hline
		$	\max epO	$&$	0.613	$&$	0.3992	$&$	0.3498	$&$	0.2442	$&$	0.2295	$&&$	0.5991	$&$	0.3994	$&$	0.3497	$&$	0.2435	$&$	0.2306	$\\	\hline
		$	\max epR	$&$	0.6124	$&$	0.3859	$&$	0.3512	$&$	0.2396	$&$	0.1895	$&&$	0.5971	$&$	0.3834	$&$	0.3511	$&$	0.2387	$&$	0.1926	$\\	
		\bottomrule[1pt]
	\end{tabular}
	\caption{Statistics for Rao index over $N\in\set{N}_{R,W,t,j}$.} 
	\label{tab_3_RWR}
\end{table}
We immediately notice that even with the greatest multilinear rank components, the average original and relative error per pixel is quite high if compared with the ones in Rényi case. We again remark that $16\%$ error per pixel on average has been obtained using only $16.4\%$ of the total information. However even in the best case the relative error per pixel is $14.5\%$, which is quite enough if compared with the minimum error per pixel in the Europe previous case.
Besides we notice that the relative error per pixel is slightly lower that the original ones as in the previous Earth case. As before we have that when the first two components of the multilinear rank are lower than $500$, ST-HOSVD techniques leads to better results than T-HOSVD. Moreover as in the previous case the variance decreases when the first two multilinear rank components grow. \\

Before moving to the analysis of error related to Rao index over $\set{N}_{N,h,p,j}$ for every $h\in\{E,W\}$, for every $p\in\{T,S\}$ and for every $j\in\{1,\dots,5\}$, we show the best case for Rao index for Europe and the worst case for Earth dataset in the following examples. In both the case we use as approximation technique T-HOSVD.
\begin{example}
	\label{ex3_3_1}
	Looking at Rao index computed over Europe NDVI of December 2014, in Figure \ref{fig3_3_1},
		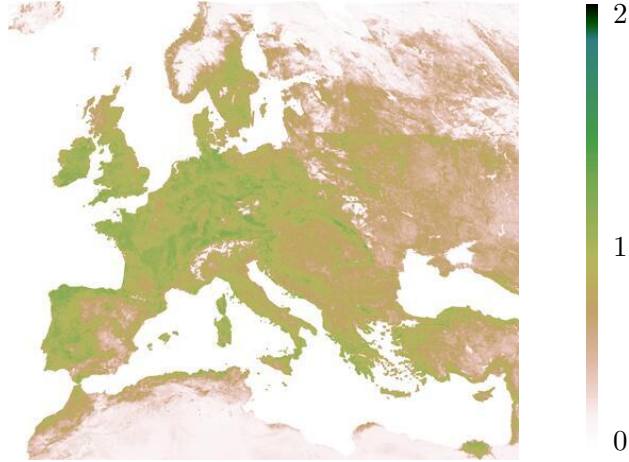
\begin{figure}
		\centering
		\subfloat{\includegraphics[scale=0.4]{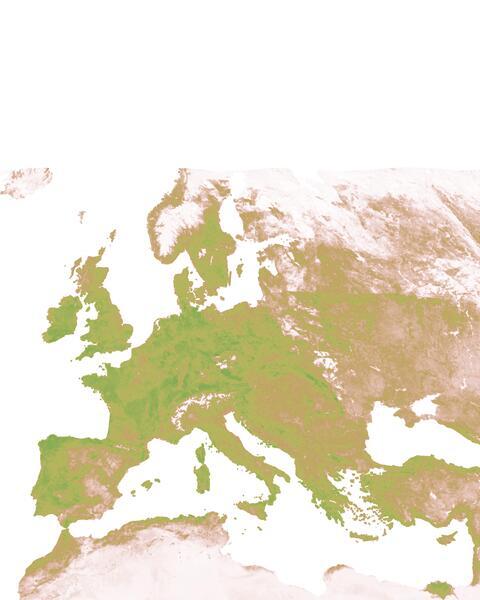}}\qquad
			\subfloat{\begin{tikzpicture}
				\node (img) {\includegraphics[scale=0.35,angle =90, origin=c]{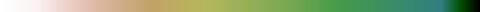}};
				\node [below right,text width=0.5cm,align=center] at (img.north){\normalcolor 2};
				\node [below right,text width=0.5cm,align=center] at (img.center){\normalcolor 1};
				\node [above right,text width=0.5cm,align=center] at (img.south){\normalcolor 0};
				\end{tikzpicture}}\\
			\caption{Rao index computed over NASA NDVI of December 2014.}
			\label{fig3_3_1}
		\end{figure}
	 we immediately notice that biodiversity is much lower than in the Rényi case, \ref{ex3_2_1}. Besides we underline that even in this case Rao index computed over self-made NDVIs is enough close to its computation over NASA NDVI. Lastly we highlight that in this particular NDVI element the north Europe area is entirely set to missing values, which are shown with white colour. Therefore we can assume that the error is the minimum since this peculiarity.\\
	Next we have in Figure \ref{fig3_3_2:a} the same index computed over self-made NDVI and the approximated Rao estimated at different multilinear rank.
		\begin{figure}
		\centering
		\subfloat[Relative approximation]{\label{fig3_3_2:a}\includegraphics[scale=0.28]{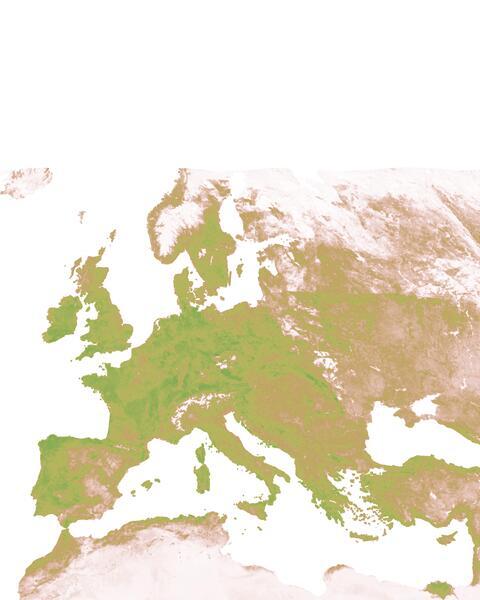}}\qquad
		\subfloat[Component rank $10$ compression]{\label{fig3_3_2:b}\includegraphics[scale=0.28]{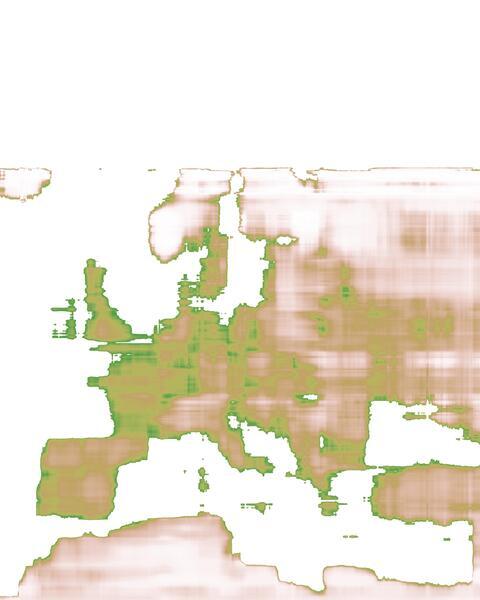}}\qquad
		\subfloat[Component rank $50$ compression]{\label{fig3_3_2:c}\includegraphics[scale=0.28]{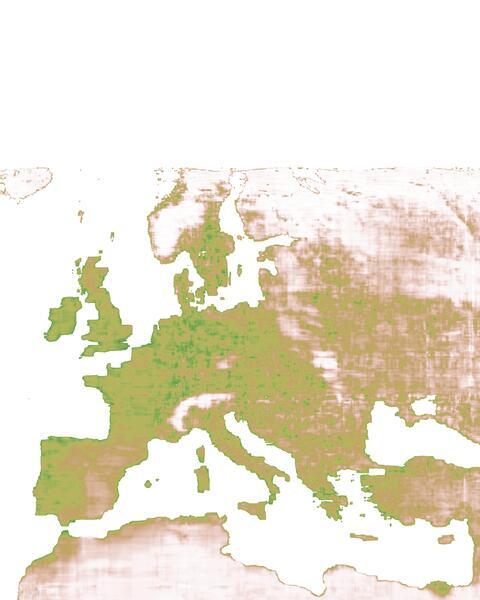}}\\
		\subfloat[Component rank $100$ compression]{\label{fig3_3_2:d}\includegraphics[scale=0.28]{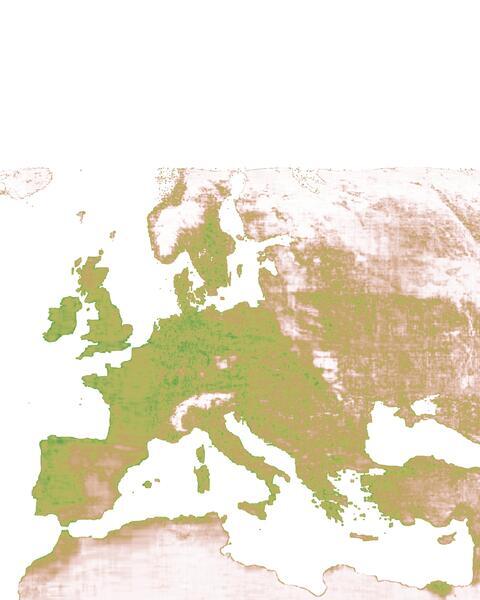}}\qquad
		\subfloat[Component rank $500$ compression]{\label{fig3_3_2:e}\includegraphics[scale=0.28]{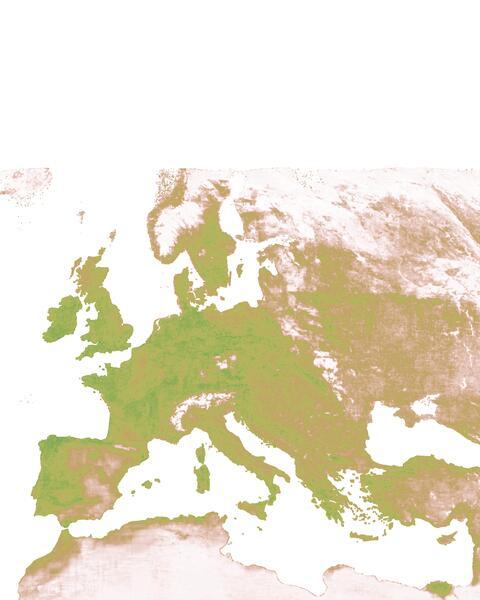}}\qquad
		\subfloat[Component rank $1000$ compression]{\label{fig3_3_2:f}\includegraphics[scale=0.28]{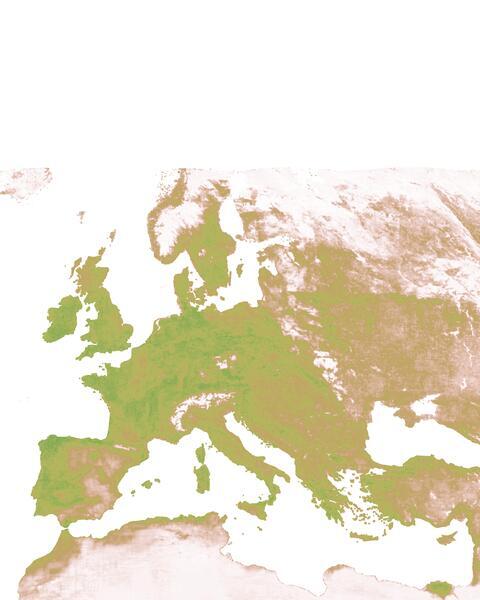}}
		\caption{Approximation of Rao index for December 2014, from NDVI of $\set{N}_{R,E,S,j}\cup\set{N}_{E}$.}
			\label{fig3_3_2}
		\end{figure} 
	It is significant that it is hard to distinguish the Rao index computed over NASA NDVI from the ones computed over the forth and the fifth approximated NDVI. 
\end{example}
Next we analyse the worst case of Rao index for the Earth dataset.
\begin{example}
	\label{ex3_3_2}
	Firstly in Figure \ref{fig3_3_4} we present Rao index over NASA NDVI of May 2014. 
	\begin{figure}
	\centering
	\subfloat{\includegraphics[scale=0.45]{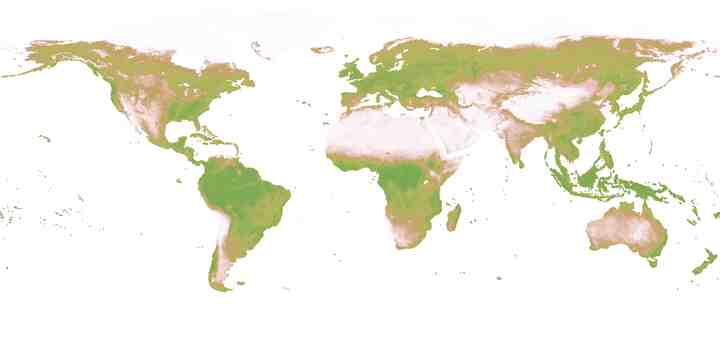}}\qquad
		\subfloat{\begin{tikzpicture}
			\node (img) {\includegraphics[scale=0.20,angle =90, origin=c]{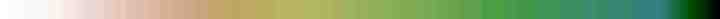}};
			\node [below right,text width=0.5cm,align=center] at (img.north){\normalcolor 2};
			\node [below right,text width=0.5cm,align=center] at (img.center){\normalcolor 1};
			\node [above right,text width=0.5cm,align=center] at (img.south){\normalcolor 0};
			\end{tikzpicture}}\\
		\caption{Rao index computed over NASA NDVI of December 2014.}
		\label{fig3_3_4}
	\end{figure}
	As in the Europe case, biodiversity estimated by Rao index has lower values. In the tropical regions for example we have the highest values, which are close to $2$. Moreover there are some areas as the Sahara desert which present extremely low biodiversity values.
	
	Next we show in Figure \ref{fig3_3_5} the Rao index computed over self-made and approximated NDVIs.
	\begin{figure}
		\centering
		\subfloat[Relative approximation]{\label{fig3_3_5:a}\includegraphics[scale=0.25]{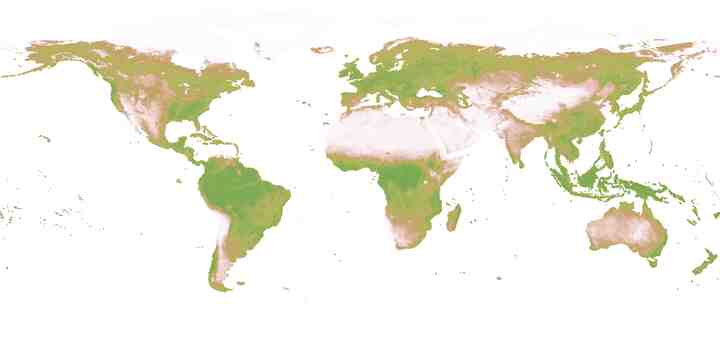}}\qquad
			\subfloat[Component rank $10$ compression]{\label{fig3_3_5:b}\includegraphics[scale=0.25]{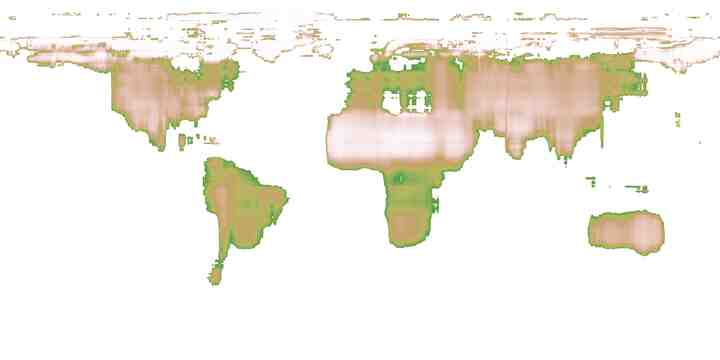}}\\
			\subfloat[Component rank $50$ compression]{\label{fig3_3_5:c}\includegraphics[scale=0.25]{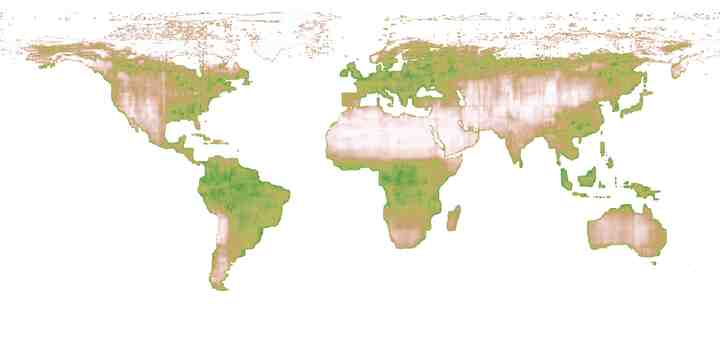}}\quad
			\subfloat[Component rank $100$ compression]{\label{fig3_3_5:d}\includegraphics[scale=0.25]{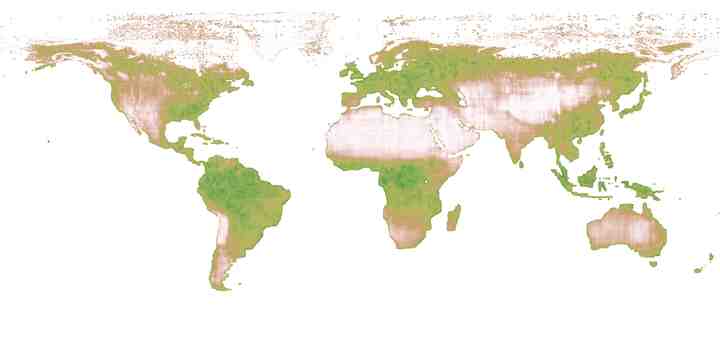}}\\
			\subfloat[Component rank $500$ compression]{\label{fig3_3_5:e}\includegraphics[scale=0.25]{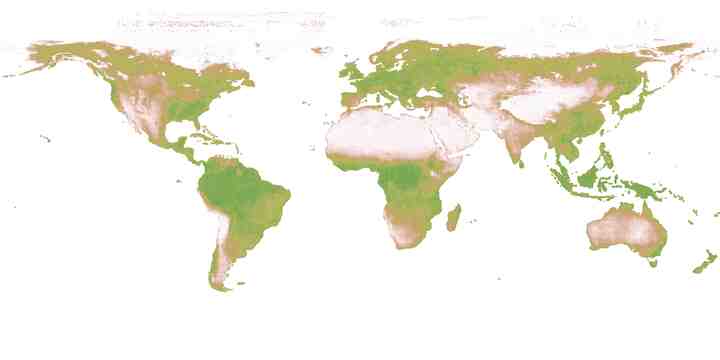}}\qquad
			\subfloat[Component rank $1000$ compression]{\label{fig3_3_5:f}\includegraphics[scale=0.25]{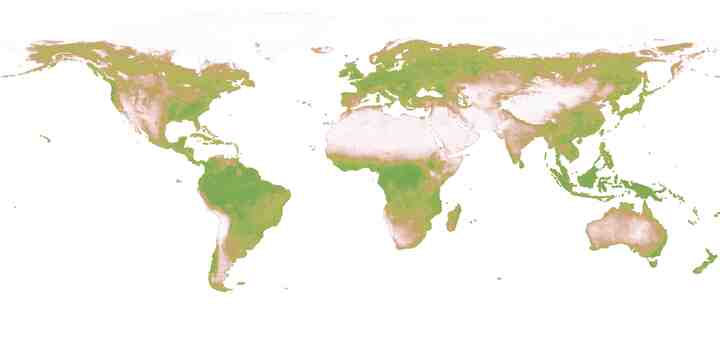}}\\
			\caption{Approximation of Rao index for May 2014, from NDVI of $\set{N}_{R,W,S,j}\cup\set{N}_{W}$.}
			\label{fig3_3_5}
		\end{figure} 
	It is clear observing the approximated Rao pictures and index of \ref{fig3_3_4} that the high error is linked with the overestimation of biodiversity in the North pole area. Indeed it is a remarkable phenomenon in the first four approximations. In Figure \ref{fig3_3_5:f} it tends to decrease, but it is still appreciable. Besides also the majority of the Pacific ocean islands are missing in all the approximations. These regions were probably related to error also in Example \ref{ex3_2_2}.
\end{example}
As before, the following step is computing Rao index over approximated NDVI of $\set{N}_{N,h,t,j}$ for every $h\in\{E,W\}$, for every $t\in\{T,S\}$ and $j\in\{1,\dots,5\}$. 
\begin{definition}
	Let $\set{R}_{N,h,t,j}$ be the set of Rao index computed over elements of $\set{N}_{N,h,t,j}$ for every $h\in\{E,W\}$, for every $t\in\{T,S\}$ and for every $j\in\{1,\dots,5\}$ and we call them $i_j$-\emph{approximated estimates} for every $i_j\in\set{R}$ and for every $j\in\{1,\dots,5\}$.
\end{definition}

Then we compute the error with respect to the original estimates, i.e.
\[\norm[2]{A_k - C_{k,j}}
\]
for every $A_k\in\set{R}_h$ and for every $C_{k,j}\in\set{R}_{N,h,T,j}\cup\set{R}_{N,h,S,j}$, for every $k\in\{1,\dots,n_h\}$, for every $j\in\{1,\dots,5\}$ and for every $h\in\{E,W\}$.
Moreover we compute the error with respect to relative estimates, i.e.
\[\norm[2]{B_k - C_{k,j}}
\]
for every $B_k\in\tilde{\set{R}}_h$ and for every $C_{k,j}\in\set{R}_{N,h,T,j}\cup\set{R}_{N,h,S,j}$, for every $k\in\{1,\dots,n_h\}$, for every $j\in\{1,\dots,5\}$ and for every $h\in\{E,W\}$.

In conclusion we report some statistics about the original and the relative errors per pixel, stored respectively in vector $epO$ and $epR$ for the Europe dataset.
\begin{table}
	\hspace*{- 6mm}
	\centering 
	\footnotesize
	\begin{tabular}{ccccccccccccc} %
		\toprule[1pt]
		& \multicolumn{5}{c}{\textbf{T-HOSVD}} &&
		\multicolumn{5}{c}{\textbf{ST-HOSVD}} \\ 
		\cmidrule{2-6} \cmidrule{8-12}
		\textbf{Rank}	&$	10	$&$	50	$&$	100	$&$	500	$&$	1000	$&&$	10	$&$	50	$&$	100	$&$	500	$&$	1000	$\\	\hline
		$	\E[epO]	$&$	0.6392	$&$	0.3641	$&$	0.2962	$&$	0.2094	$&$	0.1969	$&&$	0.6332	$&$	0.3632	$&$	0.2953	$&$	0.2111	$&$	0.1991	$\\	\hline
		$	\V[epO]	$&$	0.0049	$&$	0.0012	$&$	0.001	$&$	0.003	$&$	0.0029	$&&$	0.0044	$&$	0.0012	$&$	0.0009	$&$	0.0029	$&$	0.0029	$\\	\hline
		$	\E[epR]	$&$	0.6396	$&$	0.3647	$&$	0.2974	$&$	0.2115	$&$	0.1999	$&&$	0.6335	$&$	0.3638	$&$	0.2964	$&$	0.2131	$&$	0.2021	$\\	\hline
		$	\V[epR]	$&$	0.0048	$&$	0.0013	$&$	0.0011	$&$	0.0036	$&$	0.0043	$&&$	0.0044	$&$	0.0013	$&$	0.0011	$&$	0.0035	$&$	0.0042	$\\	\hline
		$	\min epO	$&$	0.4801	$&$	0.2741	$&$	0.2204	$&$	0.1357	$&$	0.1246	$&&$	0.4777	$&$	0.2739	$&$	0.2202	$&$	0.1378	$&$	0.1269	$\\	\hline
		$	\min epR	$&$	0.4801	$&$	0.2741	$&$	0.2204	$&$	0.1357	$&$	0.1246	$&&$	0.4778	$&$	0.2739	$&$	0.2202	$&$	0.1377	$&$	0.1269	$\\	\hline
		$	\max epO	$&$	0.7505	$&$	0.4167	$&$	0.3497	$&$	0.3933	$&$	0.3953	$&&$	0.7403	$&$	0.4155	$&$	0.3491	$&$	0.3946	$&$	0.3949	$\\	\hline
		$	\max epR	$&$	0.7504	$&$	0.4444	$&$	0.4259	$&$	0.4487	$&$	0.5458	$&&$	0.7402	$&$	0.4506	$&$	0.4193	$&$	0.4472	$&$	0.547	$\\
		\bottomrule[1pt]
	\end{tabular}
	\caption{Statistics for Rao index over $N\in\set{N}_{N,E,t,j}$.} 
	\label{tab_3_NER}
\end{table}
Comparing Table \ref{tab_3_RER} with Table \ref{tab_3_NER}, we notice that the average original and relative error per pixel are greater in the second case. In other words, starting from a tensor with repeated NIR band is not convenient for computing Rao index over Europe elements. Besides we underline that the minimum relative and original error are realised when the starting tensor has RED band repeated. We can not affirm nothing about the maximum original and relative error, since for $i_j\in\{10,100\}$ it is lower when the repeated band is the NIR, while for $i_j\in\{50, 500, 1000\}$ when it is repeated RED band. Lastly we highlight as always for the Europe dataset that the average original error is lower than the relative one. As in the repeated RED case, we remark that ST-HOSVD is convenient when the first two components of multilinear are strictly lower than $500$, otherwise T-HOSVD provides better results. 
Then we present the final statistics for original and relative errors of Rao estimates for Earth dataset. As always in vector $epO$ and in $epR$ are respectively stored the original and relative errors per pixel. 
\begin{table}[H]
	\hspace*{- 6mm}
	\centering 
	\footnotesize
	\begin{tabular}{ccccccccccccc} %
		\toprule[1pt]
		& \multicolumn{5}{c}{\textbf{T-HOSVD}} &&
		\multicolumn{5}{c}{\textbf{ST-HOSVD}} \\ 
		\cmidrule{2-6} \cmidrule{8-12}
		\textbf{Rank}	&$	10	$&$	50	$&$	100	$&$	500	$&$	1000	$&&$	10	$&$	50	$&$	100	$&$	500	$&$	1000	$\\	\hline
		$	\E[epO]	$&$	0.5366	$&$	0.3419	$&$	0.2869	$&$	0.1754	$&$	0.1636	$&&$	0.5353	$&$	0.3417	$&$	0.287	$&$	0.1761	$&$	0.1651	$\\	\hline
		$	\V[epO]	$&$	0.0023	$&$	0.0008	$&$	0.0007	$&$	0.0005	$&$	0.0003	$&&$	0.0022	$&$	0.0008	$&$	0.0007	$&$	0.0005	$&$	0.0003	$\\	\hline
		$	\E[epR]	$&$	0.535	$&$	0.3389	$&$	0.2835	$&$	0.1657	$&$	0.1525	$&&$	0.5337	$&$	0.3387	$&$	0.2836	$&$	0.1664	$&$	0.1541	$\\	\hline
		$	\V[epR]	$&$	0.0021	$&$	0.0006	$&$	0.0006	$&$	0.0003	$&$	0.0001	$&&$	0.002	$&$	0.0006	$&$	0.0006	$&$	0.0003	$&$	0.0001	$\\	\hline
		$	\min epO	$&$	0.4481	$&$	0.2873	$&$	0.2385	$&$	0.1431	$&$	0.1357	$&&$	0.4487	$&$	0.2881	$&$	0.2394	$&$	0.1437	$&$	0.1376	$\\	\hline
		$	\min epR	$&$	0.4486	$&$	0.2885	$&$	0.2385	$&$	0.1382	$&$	0.1331	$&&$	0.4483	$&$	0.2887	$&$	0.2394	$&$	0.1391	$&$	0.135	$\\	\hline
		$	\max epO	$&$	0.6063	$&$	0.3887	$&$	0.3458	$&$	0.2352	$&$	0.226	$&&$	0.6006	$&$	0.389	$&$	0.3466	$&$	0.2352	$&$	0.2267	$\\	\hline
		$	\max epR	$&$	0.6057	$&$	0.375	$&$	0.3478	$&$	0.2307	$&$	0.1746	$&&$	0.5997	$&$	0.3743	$&$	0.3487	$&$	0.2314	$&$	0.1768	$\\
		\bottomrule[1pt]
	\end{tabular}
	\caption{Statistics for Rao index over $N\in\set{N}_{N,W,t,j}$.} 
	\label{tab_3_NWR}
\end{table}
In this case comparing Table \ref{tab_3_RWR} with Table \ref{tab_3_NWR}, we notice that starting tensors with repeated NIR band provides on average better results. Indeed for both the decomposition techniques at the fifth approximation the average original and relative error per pixel is around $17.5\%$ in the first case and $16.5\%$ in the second one. This $1\%$ difference between the RED and NIR case is present also in the minimum and maximum relative and original error per pixel for both the decomposition techniques. Besides we underline that also in this case ST-HOSVD is convenient when the first two multilinear components are lower than $100$, otherwise T-HOSVD is preferable. In conclusion to these analysis we want to empathise that an average error of $16.5\%$ per pixel is not much, if we remind that we use only $16.4\%$ of the total information available.\\ 
Lastly as before, we display the Rao index computed for December 2014 and May 2014, starting from tensors with repeated NIR bands.
\begin{example}
	\label{ex3_3_3}
	Firstly we remark that December 2014 realises the minimum original and relative error, also starting from tensors with repeated NIR band. 
	In Figure \ref{fig3_3_3} there are approximated Rao estimates at different multilinear rank, for this NIR repeated case.
		\begin{figure}
			\centering
			\subfloat[Component rank $10$ compression]{\label{fig3_3_3:b}\includegraphics[scale=0.28]{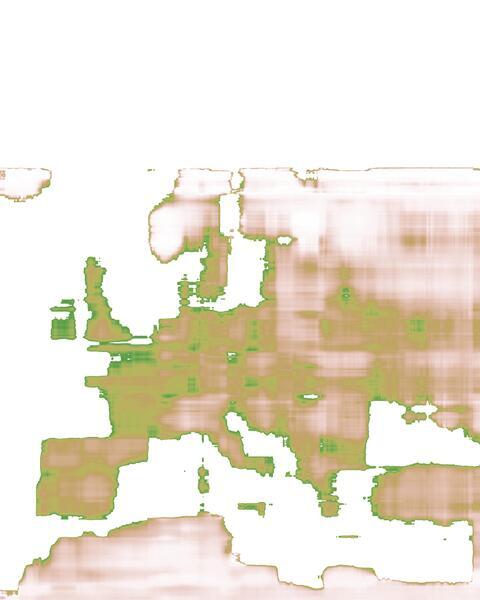}}\quad
			\subfloat[Component rank $50$ compression]{\label{fig3_3_3:c}\includegraphics[scale=0.28]{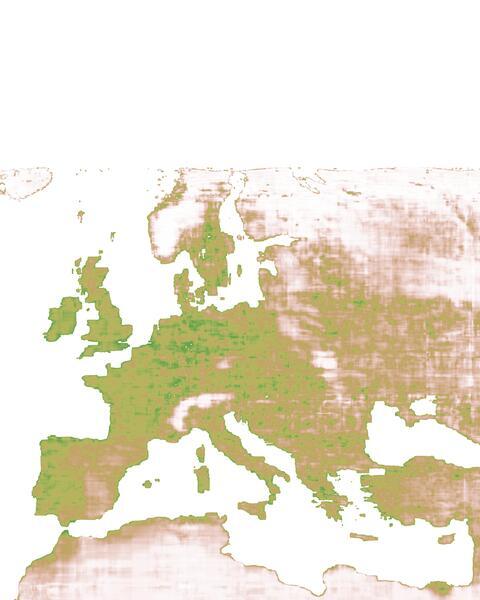}}\quad
			\subfloat[Component rank $100$ compression]{\label{fig3_3_3:d}\includegraphics[scale=0.28]{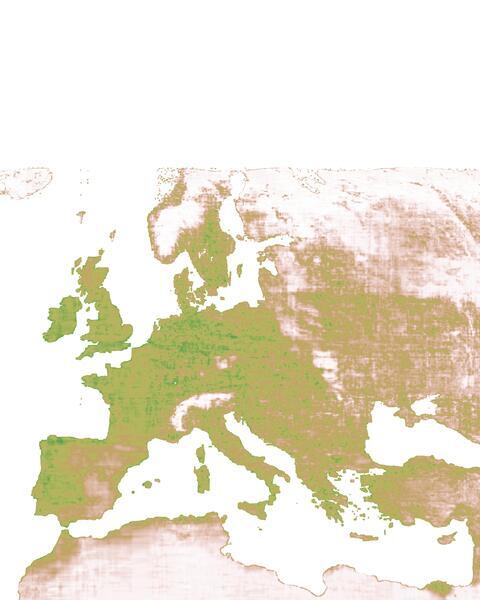}}\\
			\subfloat[Component rank $500$ compression]{\label{fig3_3_3:e}\includegraphics[scale=0.28]{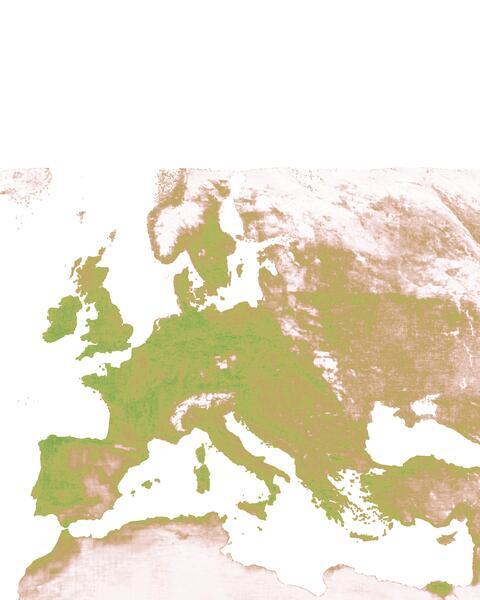}}\%
			\subfloat[Component rank $1000$ compression]{\label{fig3_3_3:f}\includegraphics[scale=0.28]{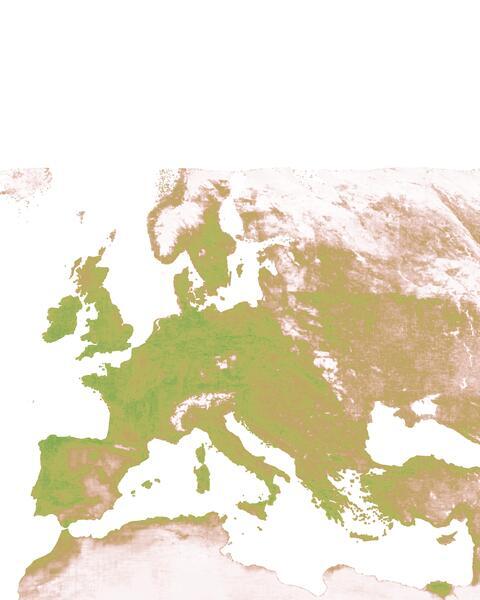}}\qquad
			\subfloat{\begin{tikzpicture}
				\node (img) {\includegraphics[scale=0.25,angle =90, origin=c]{MigliorRaoERaoL.jpg}};
				\node [below right,text width=0.5cm,align=center] at (img.north){\normalcolor 2};
				\node [below right,text width=0.5cm,align=center] at (img.center){\normalcolor 1};
				\node [above right,text width=0.5cm,align=center] at (img.south){\normalcolor 0};
				\end{tikzpicture}}
			\caption{Approximation of Rao index for December 2014, from NDVI of $\set{N}_{N,E,S,j}$.}
			\label{fig3_3_3}
		\end{figure} 
	As before, our eyes hardly perceive differences between Rao computed over NASA or self-made NDVI and over the last three approximated NDVIs.
\end{example}
\begin{example}
	\label{ex3_5_4}
	Firstly we remark that May 2014 do not realise the minimum original and relative error starting from tensors with repeated NIR band. Therefore these are not the worst approximations in the Earth dataset, in this second case. 
	In Figure \ref{fig3_3_6} there are approximated Rao estimates at different multilinear rank, for this NIR repeated case.
		\begin{figure}
			\centering
			\subfloat[Component rank $10$ compression]{\label{fig3_3_6:b}\includegraphics[scale=0.25]{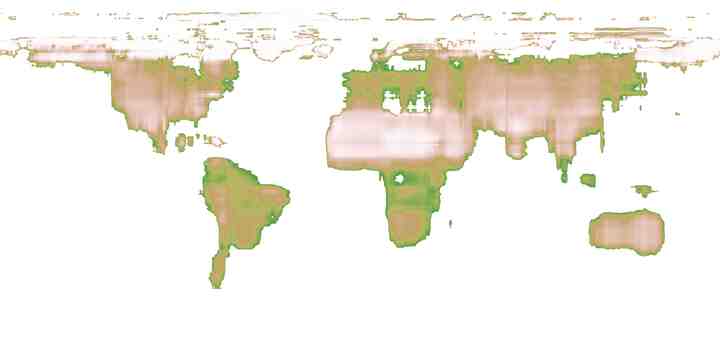}}\quad
			\subfloat[Component rank $50$ compression]{\label{fig3_3_6:c}\includegraphics[scale=0.25]{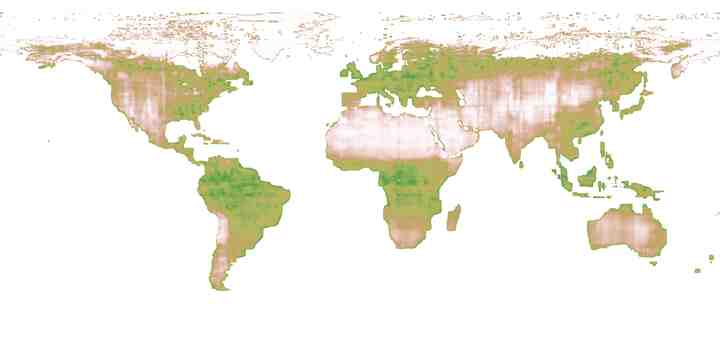}}\\
			\subfloat[Component rank $100$ compression]{\label{fig3_3_6:d}\includegraphics[scale=0.25]{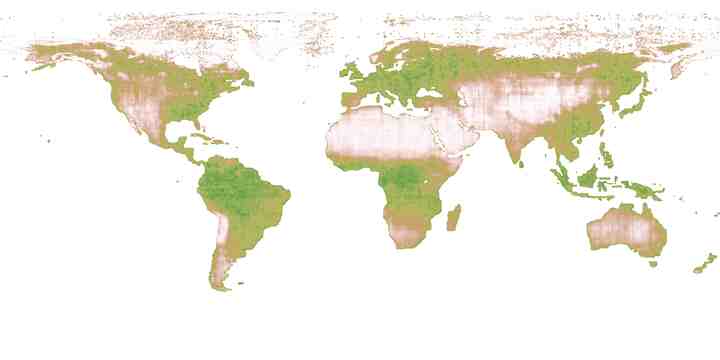}}\qquad
			\subfloat[Component rank $500$ compression]{\label{fig3_3_6:e}\includegraphics[scale=0.25]{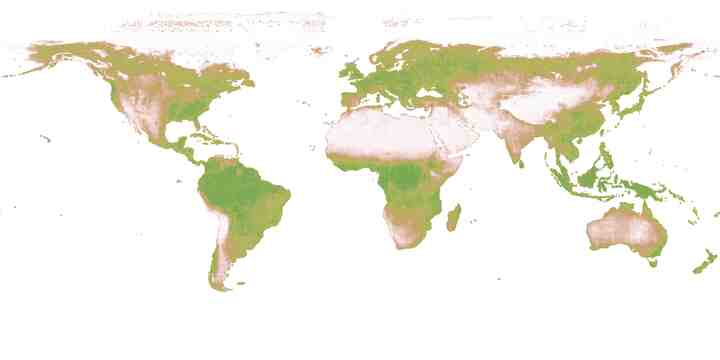}}\\
			\subfloat[Component rank $1000$ compression]{\label{fig3_3_6:f}\includegraphics[scale=0.45]{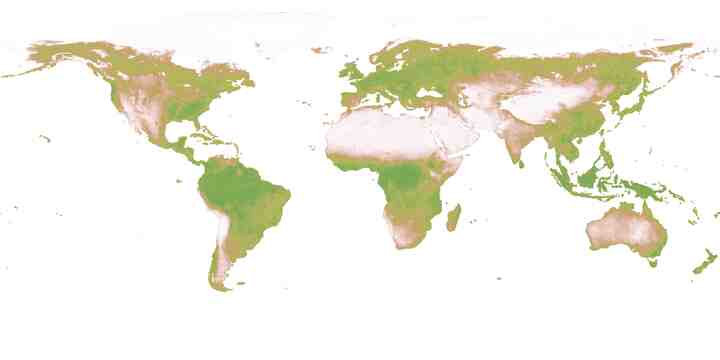}}\qquad
			\subfloat{\begin{tikzpicture}
				\node (img) {\includegraphics[scale=0.2,angle =90, origin=c]{PeggiorRaoWRaoL.jpg}};
				\node [below right,text width=0.5cm,align=center] at (img.north){\normalcolor 2};
				\node [below right,text width=0.5cm,align=center] at (img.center){\normalcolor 1};
				\node [above right,text width=0.5cm,align=center] at (img.south){\normalcolor 0};
				\end{tikzpicture}}
			\caption{Approximation of Rao index for May 2014, from NDVI of $\set{N}_{N,W,S,j}$.}
			\label{fig3_3_6}
		\end{figure} 
	As in Figure \ref{fig3_3_5} the problematic areas are the North pole and the Pacific ocean. Indeed we notice that there is a biodiversity overestimation in the Arctic regions, which decreases for increasing multilinear rank components. Moreover even in the fifth approximation there are missing island in the Pacific ocean. However we can not easily notice much difference with respect to Figure \ref{fig3_3_5}. 
\end{example}
\section{Conclusion}
We have shown that the presented approach is extremely convenient for Rényi index estimation to save storage memory. As reported in Tables \ref{tab_3_REI},\ref{tab_3_NEI}, \ref{tab_3_RWI} and \ref{tab_3_NWI} the average error per pixel is around $5.5\%$ for Earth dataset and is close to $7.6\%$ for the Europe's one when using respectively $16.4\%$ and $14.8\%$ of the total tensor information. Moreover we have seen that starting from tensor with repeated NIR bands is more convenient than with RED repeated, in the Rényi case. For Rao case starting with twice NIR rasters is more convenient only for Earth dataset. Lastly we have remarked that in Rényi case T-HOSVD makes on biodiversity estimation lower error than ST-HOSVD when the first two multilinear rank components are relatively low.\par
In the Rao case we face greater on average original and relative error per pixel. Indeed in Tables \ref{tab_3_RER} and \ref{tab_3_RWR} we have an average error per pixel close to $19.5\%$ and $17\%$ for Europe and for Earth dataset respectively. In this case we have noticed that the roles of T-HOSVD and ST-HOSVD are inverted. Indeed T-HOSVD makes on biodiversity estimation lower error than ST-HOSVD with the first two multilinear rank components are relatively great.\par
In conclusion we believe that the presented work might help ecologists in their remote sensed plant biodiversity estimation. Indeed fixed a certain accuracy, they can compress through T-HOSVD and ST-HOSVD the tensors with NIR and RED bands, to save storage memory and at the same time computing with fixed tolerance the biodiversity estimates. \\
Lastly if it was possible to directly download decomposed tensors, ecologists would benefit of a significant computer memory saving.
\bibliographystyle{amsalpha}
\begin{small}
\bibliography{Art}
\end{small}

\section{Appendix}
\subsection{Rao's and Rényi's codes}\label{app:1}
Remind that for every raster of size $(m,n)$, fixed $l$ the side of the moving window, the biodiversity index is computed $l\times m\times n$ times. To speed up the entire work we decide to implement a parallel version of the biodiversity index algorithm. When a computer executes a parallel functions its cores perform independently different tasks at the same time. In our case we want that each core of the used machine works on different position of the moving window. Consequently to implement this mechanism we needed a parallel computing library compatible with \texttt{Python}, the chosen programming language. We preferred \texttt{Joblib}, cf.~\cite{res4}, since its ease of use. Other two used libraries are \texttt{itertools}, to create iterable elements and \texttt{spatial} from \texttt{SciPy}, to compute the distance element wise for two matrices. The main Rao computation code is
\begin{lstlisting}[language=Python]
#### computation
import numpy as np
import scipy
from scipy import spatial
import itertools
#### parallelisation 
import joblib
from joblib import Parallel, delayed
import multiprocessing

def raop(rw):
	def raout (cl, rw = rw, rasterm = rasterm, missing = missing, w = w, distance_m = distance_m):
		tw_labels, tw_values = np.unique(rasterm[(rw-w):(rw+w+1),(cl-w):(cl+w+1)], return_counts=True)
		if len(np.where(tw_labels == missing)) != 0:
			tw_values = np.delete( tw_values, np.where(tw_labels == missing))
			tw_labels = np.delete( tw_labels, np.where(tw_labels == missing))
		if len(tw_values) > 1:
			d1 = spatial.distance.cdist(np.diag(tw_labels), np.diag(tw_labels), distance_m)
			p = tw_values/np.sum(tw_values)
			p1 = np.zeros((len(tw_values),len(tw_values)))
			comb = np.array([x[0]*x[1] for x in list(itertools.combinations(p, 2))])
			p1[np.triu_indices(len(tw_values), k=1)] = comb
			p1[np.tril_indices(len(tw_values), k=-1)] = comb
			return ((np.sum(np.multiply(p1,d1))))
		elif len(tw_values) == 1:
			return (((0)))
		else:
			return ((missing))
	Raout = Parallel(n_jobs = NcCores)(delayed(raout)(cl) for cl in range(w,c-w))
	return (Raout)
out[w:(r-w), w:(c-w)] = (np.asarray(Parallel(n_jobs = NcCores)(delayed(raop)(rw) for rw in range(w,(r-w)))).reshape(r-2*w,c-2*w))
\end{lstlisting} 
Since the moving window scrolls over rows and over columns, we define two parallelised functions one inside the others, \texttt{raop} whose variable is just the row index and \texttt{raout}. This second function takes as variable the column index and has other set parameters: \texttt{rasterm} the raster, \texttt{rw} the row index, \texttt{missing} a value used in the raster when data are not present, \texttt{w} the window side and a function \verb|distance_m| to compute the distance between raster values. The first step is storing the values and their frequencies of the raster area covered by the moving window in the arrays \verb|tw_labels| and \verb|tw_values| respectively. Then we check if there is the \texttt{missing} value in the considered area: if it is present, we remove it and its frequency from the storing arrays. Next if there are at least two different elements in the \verb|tw_values| array, we compute their distance with the function \texttt{spatial.distance.cdist} and save the result in the \texttt{d1} matrix. In \texttt{p} we put the relative frequencies, obtained from the absolute ones. We define a $0$ matrix \texttt{p1}, with the same size of \texttt{d1}. We compute and store in \texttt{comb} all the possible combination between different elements of vector \texttt{p}. We assign these values to the upper and lower triangular part of the matrix \texttt{p1}. Lastly we return the sum of the elements of the product matrix between \texttt{d1} and \texttt{p1}. Notice that the product matrix will coherently have diagonal null elements, since the distance between a pixel value and itself is $0$.\\ If there is just one element in vector \verb|tw_values|, we return $0$ since the difference between a pixel value and itself is $0$. Otherwise, if \verb|tw_values| is an empty vector, we return \texttt{missing}. Inside function \texttt{raop} we define function \texttt{raout} and we do the first \texttt{Parallel} call. Outside it we do the second \texttt{Parallel call}. Notice that this function is thought to compute Rao's index when the moving window is completely contained into the raster. When this condition is not satisfied, to speed up computation, there is a special implementation of Rao's, available here~\cite{res5}.
With a similar approach, we implemented Rényi's index, whose core is
\begin{lstlisting}[language=Python]
#### computation
import numpy as np
import scipy
#### parallelisation 
import joblib
from joblib import Parallel, delayed
import multiprocessing

def IndexOP (rw):
	def IndexOut (cl, rw = rw, rasterm = rasterm, missing = missing, w = w, alpha=alpha, base = base):
		tw_labels, tw_values = np.unique(rasterm[(rw-w):(rw+w+1),(cl-w):(cl+w+1)], return_counts=True)
		if len(np.where(tw_labels == missing)[0]) != 0:
			tw_values = np.delete( tw_values, np.where(tw_labels == missing))
		if len(tw_values) != 0:
			p = tw_values/np.sum(tw_values)
			if np.log(np.sum(p**alpha)) == 0:
				return(0)
			else:
				return((1/(1-alpha)) * np.log(np.sum(p**alpha)) / np.log(base))
		else:
			return (missing)
	Index_Out = Parallel(n_jobs = NcCores)(delayed(IndexOut)(cl) for cl in range(w,c-w))
	return (Index_Out)
out[w:(r-w), w:(c-w)] = np.asarray(Parallel(n_jobs = NcCores)(delayed(IndexOP)(rw) for rw in range(w,r-w))).reshape(r-2*w,c-2*w)
\end{lstlisting} 
Similarly we declare function \texttt{IndexOP}, with just row index as variable and inside it function \texttt{IndexOut}. This one takes as parameters some equals to \texttt{raout}'s ones. The variable proper of \texttt{IndexOut} are \texttt{alpha}, the Rényi's parameter, and \texttt{base}, the logarithm base. The code structure is similar to \texttt{raout}. We compute and store the values, present inside the raster area covered by the moving window, and their frequencies. We check and in case delete \texttt{missing} value and its frequency. If there is still an element into \verb|tw_values|, we compute the relative frequencies and return the Rényi index. We test if the index final value is $0$ and return $0$ to avoid sign problem. If \verb|tw_values| is an empty vector, we return \texttt{missing} value. We do a first \texttt{Parallel} call of \texttt{IndexOut} inside \texttt{IndexOP} and a second one of \texttt{IndexOP} outside it. Even in this case the presented code will work only when the moving window is entirely contained in the raster. To speed up the computations, we develop a special version of this code which works when the moving window is not fully contained. It is available here ~\cite{res5}.
 
\subsection{Approximation codes for T-HOSVD}\label{app:code:mat_THOSVD} 
In order to implement the approximation Algorithm \ref{T-HOSVD_algo} of T-HOSVD we need a python library expressly developed to manage tensors. Moreover we also want that this library to interact properly with \texttt{NumPy} the python most used library for numerical computation, cf.~\cite{res1}. Therefore we choose \texttt{TensorLy}, cf.~\cite{res2}. This recently developed library, firstly presented in 2016, wants to make tensor study and manipulation easy and accessible. Besides their creators projected \texttt{TensorLy} to perfectly match with other famous python libraries, as \texttt{NumPy}, \texttt{SciPy}. They developed most of the main tensor operations and related functions. To compute the singular value decomposition of tensors flattening, the central step in T-HOSVD and ST-HOSVD algorithm, we need a python function able to manage huge arrays. This is not a trivial task, since development of python functions is left to singular initiative. We decide to use \texttt{svds} function from \texttt{SciPy} sparse linear algebra function, cf.~\cite{res3}. This implementation of SVD takes advantage of matrix sparsity in performing the matrix-vector multiplication. The final implementation of T-HOSVD is

\begin{lstlisting}[language=Python]
import numpy as np
import tensorly as tl
from tensorly import decomposition as decompose
from tensorly import tenalg as Tla
import scipy
from scipy import spatial
from scipy.sparse.linalg import svds

def T_hosvd(T, rango, projector = Ture):
	L = []
	dim = T.shape
	for i in range(len(dim)):
		flat = tl.unfold(T,i)
		res = svds(flat, k = rango[i])
		L.append(np.transpose(res[0][0:dim[i], :]))
	core = Tla.multi_mode_dot(T,L)
	if projector:
		P = [np.transpose(u) for u in L]
		return [core,P]
	return core
\end{lstlisting}
This function takes as input variable a tensor \texttt{T} and a list or a tuple, whose values are the target multilinear rank components. 
Inside the \verb|T_hosvd| we declare an empty list, \texttt{L}, in which at the $i$-th step we store matrix \verb|U_i| from the thin SVD of the $i$-th flattening for every $i\in\{1,\dots,d\}$. The \texttt{dim} variable is a tuple containing the size of tensor \texttt{T}. Then for each direction we compute the flattening and its SVD. After the for loop, we get the core tensor with the multilinear product between a list of matrices, our projectors, and the original tensor \texttt{T}.
Notice that there is the option \texttt{projector}, with \texttt{True} as default value, to return a list \texttt{L} containing the projectors matrices together with the standard result, i.e. the core tensor.
\subsection{Approximation codes for ST-HOSVD}\label{approx_STHOSVD}
The ST-HOSVD implementation of Algorithm \ref{algoSTHOSVD} is 
\begin{lstlisting}[language=Python]
import numpy as np
import tensorly as tl
from tensorly import decomposition as decompose
from tensorly import tenalg as Tla
import scipy
from scipy import spatial
from scipy.sparse.linalg import svds

def ST_hosvd(T, rango, projector = True):
	dim = T.shape
	core = T
	if projector: 
		P=[]
	for i in range(len(dim)):
		flat = tl.unfold(core,i)
		res = svds(flat, k=rango[i])
		core = Tla.mode_dot(core,np.transpose(res[0][0:dim[i],:]),mode=i)
		if projector: 
			P.append(res[0][0:dim[i],:])
	if projector:
		return [core,P]
	return core
\end{lstlisting}
The input arguments of ST-HOSVD and T-HOSVD implementations are the same. The first difference with \verb|T_hosvd| is the initial declaration of \texttt{core} tensor, set equal to \texttt{T}. Next we highlight that only if the \texttt{projector} variable is \texttt{True}, we initialize an empty list to store the projectors matrices. Another difference is the computation of the core tensor inside the for loop with the component wise product between matrix and tensor, fixed a certain direction. The basic output is still the final core tensor.

\end{document}